%% file: MixofMix-ARXIV.tex
\documentclass[12pt,letter]{article}
\usepackage{amsmath}
\usepackage{graphicx,psfrag,epsf,color}
\usepackage{enumerate}
\usepackage{natbib}
\usepackage{url} % not crucial - just used below for the URL
\usepackage{enumerate}
\usepackage{subfigure}
\usepackage{slashbox,booktabs}

\newcommand{\blind}{0}
\include{newcommands}
\renewcommand{\r}{\color{black}}
\newcommand{\bl}{\color{black}}
\usepackage[usenames,dvipsnames]{xcolor}
\newcommand{\comment}[1]{#1}%{\textcolor{Green}{#1}}
\newcommand{\commentG}[1]{\textcolor{Black}{#1}} % new changes by Sylvia

\newcommand{\KL}{$K$$\cdot$$L$}
\newcommand{\KLfac}{$(K$$\cdot$$L)!$}

\let\proglang=\textsf
\newcommand{\pkg}[1]{{\fontseries{b}\selectfont #1}}

\begin{document}

\def\spacingset#1{\renewcommand{\baselinestretch}%
{#1}\small\normalsize} \spacingset{1}

%%%%%%%%%%%%%%%%%%%%%%%%%%%%%%%%%%%%%%%%%%%%%%%%%%%%%%%%%%%%%%%%%%%%%%%%%%%%%%

\if0\blind
{
  \title{\bf Identifying Mixtures of  Mixtures Using Bayesian
  Estimation}
  \author{Gertraud Malsiner-Walli\\
    Department of Applied Statistics, Johannes Kepler University Linz\\
    and \\
    Sylvia Fr\"uhwirth-Schnatter \\
    Institute of Statistics and Mathematics, Wirtschaftsuniversit\"at Wien\\
    and \\
    Bettina Gr\"un\thanks{
    The author gratefully acknowledges support by the Austrian Science Fund (FWF): V170-N18.}\hspace{.2cm}\\
    Department of Applied Statistics, Johannes Kepler University Linz\\}
  \maketitle
} \fi

\if1\blind
{
  \bigskip
  \bigskip
  \bigskip
  \begin{center}
    {\LARGE\bf Identifying Mixtures of  Mixtures Using Bayesian
  Estimation}
\end{center}
  \medskip
} \fi

\bigskip
\begin{abstract}
  The use of a finite mixture of normal distributions in model-based
  clustering allows to capture non-Gaussian data clusters. However,
  identifying the clusters from the normal components is challenging
  and in general either achieved by imposing constraints on the model
  or by using post-processing procedures.

  Within the Bayesian framework we propose a different approach based
  on sparse finite mixtures to achieve identifiability.  We specify a
  hierarchical prior where the hyperparameters are carefully selected
  such that they are reflective of the cluster structure aimed at. In
  addition, this prior allows to estimate the model using standard MCMC
  sampling methods.  In combination with a post-processing approach
  which resolves the label switching issue and results in an
  identified model, our approach allows to simultaneously (1)
  determine the number of clusters, (2) flexibly approximate the
  cluster distributions in a semi-parametric way using finite mixtures
  of normals and (3) identify cluster-specific parameters and classify
  observations. The proposed approach is illustrated in two simulation
  studies and on benchmark data sets.

  %  Within the Bayesian framework we show that identification of
%  clusters is possible by choosing a suitable prior which
%  reflects in a principled way the information on the cluster shapes
%  aimed at. This is essential in order to be able to combine normal
%  components to form a cluster in an automatic way. Based on these
%  priors MCMC sampling can be used for model estimation. Using one run
%  of MCMC sampling the number of clusters is selected and cluster
%  distributions are automatically formed by a mixture of
%  normals. After model identification on the cluster level the
%  cluster-specific distributions can be investigated and the model
%  used for classifying observations.  The proposed approach is
%  illustrated in two simulation studies and on benchmark data sets.
\end{abstract}

\noindent%
\vfill

\newpage
\spacingset{1.45} % DON'T change the spacing!

\section{Introduction}\label{sec:intro}

In many areas of applied statistics like economics, finance or public
health it is often desirable to find groups of similar objects in a
data set through the use of clustering techniques.  % Popular heuristic
% clustering techniques such as $k$-means \citep{mac:som} are based on
% distance measures and do not easily allow to incorporate specific
% knowledge about the underlying data generating distribution.
A flexible approach to clustering data is based on  mixture
models, whereby the data in each mixture component are assumed to
follow a parametric distribution with component-specific parameters
varying over the components.
% (see \citealp{mcl-pee:fin}, and
%\citealp{Mix:Fruehwirth2006}, for a comprehensive survey of mixture
%models and their various applications).
This so-called model-based
clustering approach \citep{Mix:FraleyRaftery2002} is based on the
notion that the component densities can be regarded as the \lq\lq
prototype shape of clusters to look for\rq\rq\ \citep{Mix:Hennig2010}
and each mixture component may be interpreted as a distinct data
cluster.

Most commonly, a finite mixture model with Gaussian component
densities is fitted to the data to identify homogeneous data
clusters within a heterogeneous population. However, assuming such a
simple parametric form for the component densities implies a strong
assumption about the shape of the clusters and may lead to overfitting
the number of clusters as well as a poor classification, if not
supported by the data.  Hence, a major limitation of Gaussian mixtures
in the context of model-based clustering results from the presence of
non-Gaussian data clusters, as typically encountered in practical
applications. % such as genetics ADD.

Recent research demonstrates the usefulness of mixtures of parametric non-Gaussian
component  \commentG{densities} %distributions
such as the skew normal or skew-$t$
distribution to capture non-Gaussian data clusters, see
\cite{Mix:FruehwirthPyne2010}, \cite{Mix:LeeMcLachlan2014} and
\cite{Mix:VrbikMcNicholas2012},  \commentG{among  others}.
  However, as stated in \cite{Mix:Li2005}, for many applications it is difficult
to decide which parametric distribution is appropriate to characterize
a data cluster, especially in higher dimensions.  In addition, the
shape of the cluster densities can be of a form which is
\commentG{not easily captured} %d difficult to   escribe  accurately
by a parametric distribution. To better accommodate such data,
recent advances in model-based clustering focused on designing mixture models
with more flexible, not necessarily parametric cluster densities.

A rather appealing approach, known as mixture of mixtures, models the
non-Gaussian cluster distributions themselves by Gaussian mixtures,
exploiting the ability of normal mixtures to accurately approximate a
wide class of probability distributions.
%\citep{fer:bay_den}. %\citep{fer:bay_den,mar-wan:exa}.
Compared to a mixture with Gaussian components, mixture of mixtures
models impose a two-level hierarchical structure which is particularly
appealing in a clustering context. On the higher level, Gaussian
components are grouped together to form non-Gaussian cluster
distributions which are used for clustering the data. The individual
Gaussian component densities appearing on the lower level of the model
influence the clustering procedure only indirectly by accommodating
possibly non-Gaussian, but otherwise homogeneous cluster distributions
in a semi-parametric way.  This powerful and very flexible approach
has been employed in various ways, both within the framework of finite
and infinite mixtures.

Statistical inference for finite mixtures is generally not easy due to
problems such as label switching, spurious modes and unboundedness of
the mixture likelihood \citep[see
e.g.][Chapter~2]{Mix:Fruehwirth2006}, but estimation of a mixture of
mixtures model is particularly challenging due to additional
identifiability issues.  Since exchanging subcomponents between
clusters on the lower level leads to different cluster distributions,
while the density of the higher level mixture distribution remains the
same, a mixture of mixtures model is not identifiable from the mixture
likelihood in the absence of additional information.  {\r For example,
  strong identifiability constraints on the locations and the
  covariance matrices of the Gaussian components were imposed by
  \citet{Mix:Bartolucci2005} for univariate data and by
  \citet{Mix:DizioGuarneraRocci2007} for multivariate data to estimate
  finite mixtures of Gaussian mixtures.}

A different strand of literature pursues the idea of creating
meaningful clusters after having fitted a standard Gaussian mixture
\commentG{model} to the data.  The clusters are determined by successively
merging components according to some criterion, e.g.~the closeness of
the means \citep{Mix:Li2005}, the modality of the obtained mixture
density \citep{Chan2008,Mix:Hennig2010}, the degree of overlapping
measured by misclassification probabilities \citep{Mix:Melnykov2014}
or the entropy of the resulting partition
\citep{Mix:BaudryRafteryCeleuxLoGottardo2010}. However,
\commentG{such two-step  approaches might miss the  general cluster structure,}
%the  general cluster structure might be missed by such two-step  approaches,
see Appendix~E for an example.

% Within the Bayesian community, in the influential work by
% \citet{Chan2008}, which was subsequently extended to generate
% complex mixture structures for frontier research in molecular
% biology and biotechnology \citep{Chan2010,Cron2013,Lin2013},
% identification of the the clusters is performed by merging normal
% components with a common mode.

%  Gaussian mixture ofmixtures were introduced by the influential work
%  of \citet{Chan2008} where identification of the clusters is
%  performed by merging normal components with a common mode, and,
%  subsequently, extended to generate complex mixture structures for
%  frontier research in molecular biology and biotechnology
%  \citep{Chan2010,Cron2013,Lin2013}.
%
% Compared to the hierarchical mixture of mixtures approach proposed in this paper,

In the present paper, we identify the mixture of mixtures model within
a Bayesian framework through a hierarchical prior construction and propose
a method to simultaneously select a suitable number of clusters. In our
approach both the identification of the model and the estimation of the number of clusters is achieved by employing a
selectively informative prior parameter setting on the model parameters.

%In the present paper, we identify the mixture of mixtures model within
%a Bayesian framework through a hierarchical prior construction using a
%selectively informative prior parameter setting on the model
%parameters and propose a method to simultaneously select a suitable
%number of clusters.

%The choice of our prior parameters is derived based on
\commentG{Our  choice of  prior parameters is driven by}
 assumptions on the
cluster shapes assumed to be present in the data, thus being in line
with \citet{Mix:Hennig2010} who emphasizes that, \emph{\lq\lq it
  rather has to be decided by the statistician under which conditions
  different Gaussian mixture components should be regarded as a common
  cluster\rq\rq }.  This prior specification introduces dependence among the
subcomponent densities within each cluster, by pulling the
subcomponent means on the lower level toward the cluster center,
making the cluster distributions themselves dense and connected.  On
the higher level, the prior is based on the notion that the cluster
centers are quite distinct from each other compared to the spread of
the clusters. The choice of the hyperparameters of
this hierarchical prior turns out to be crucial in achieving
identification and \commentG{is} guided by a variance decomposition of the
data.

Regarding the estimation of the number of clusters, a sparse
hierarchical mixture of mixtures model is derived as an extension of
the sparse finite mixture model introduced in
\cite{Mix:MalsinerWalliFruehwirthGruen2014}.  There, based on
theoretical results derived by \cite{Mix:RousseauMengerson2011}, an
overfitting Gaussian mixture \comment{with $K$ components} is
specified where a sparse prior on the mixture weights has the
effect %\comment{emptying superfluous components} during MCMC sampling, by
\commentG{of assigning the observations to fewer than $K$} {\r
  components}. Thus, the number of {\r clusters} can be estimated by
the most frequent number of non-empty components {\r encountered
  during Markov chain Monte Carlo (MCMC) sampling}. In this paper, rather than using a single
multivariate Gaussian distribution, we model the component densities
in a semi-parametric way through a Gaussian mixture distribution, and
again use a sparse prior on the cluster weights to automatically
select a suitable number of clusters \comment{on the upper level}.

\commentG{Specifying a sparse prior on the weights is closely related
  to Bayesian nonparametric (BNP) Gaussian mixture models such as
  Dirichlet process mixtures
  \citep[DPMs;][]{fer:bay_den,Mix:Escobar+West:1995}.  The sparse
  prior on the cluster weights induces clustering of the observations,
  similar as for DPM{\r s} which have been applied in a clustering
  context by \cite{qui-igl:bay}, \citet{Mix:Medvedovic2004} and
  \citet{Mix:Dahl2006}, among others.}  \commentG{The hierarchical
  mixture of {\bl mixtures} model we introduce is {\bl similar  
  to} %truncated DPMs \citep{Suchard2010},
  hierarchical BNP approaches such as the hierarchical DPM
  \citep{Mix:Teh2006}.
%  truncated  DPM \citep{Suchard2010}.
  Very closely related BNP approaches are infinite mixtures of
  infinite Gaussian densities such as the nested DPM
  \citep{Mix:RodriguezDunsonGelfand2008}, the infinite mixture of
  infinite Gaussian mixtures \citep{Mix:Yerebakan2014}, and species
  mixture models \citep{Mix:ArgientoCremaschiGuglielmi2013} which
  directly work on the partition of the data.  We discuss in
  Sections~\ref{BNP} and \ref{sec:clust} similarities as well as
  differences between our approach and BNP models.}

We finally note that the  implementation effort  to
estimate our model is
moderate %.  Since we are staying within the framework of finite
% mixtures of normals and conditionally conjugate priors,
and standard MCMC methods based on data augmentation and Gibbs
sampling \citep[see][]{Mix:Fruehwirth2006} can be used.  Several
approaches proposed in the literature can be used to post-process the
MCMC draws in order to obtain a clustering of the data and also to
allow for cluster-specific inference. For our simulation studies and
applications we adapt and extend the method suggested by
\cite{Mix:Fruehwirth2006, Mix:Fruehwirth2011} which determines a
unique labeling for the MCMC draws by clustering the draws in the
point process representation.
  % and of  basing any subsequent inference on the relabeled draws.

The rest of the article is organized as follows.
Section~\ref{sec:model-specification} describes the proposed strategy,
including detailed prior specifications, and relates our method  to the two-layer BNP approaches
in  \citet{Mix:RodriguezDunsonGelfand2008} and  \citet{Mix:Yerebakan2014}.  \commentG{Clustering  and model
estimation issues} are discussed in
Section~\ref{sec:modelEstimation}.   \commentG{The performance of the proposed
strategy is evaluated in Section~\ref{sec:applications} for various  benchmark  data sets.} Section~\ref{sec:discussion} concludes.

% -------------------------------------------------------------------------------------------
%$\S\S\S\S\S\S\S\S\S\S\S\S\S\S\S$ ADD Material $ \S\S\S\S\S\S\S\S\S\S\S\S\S\S\S\S\S\S\S\S$
%
%\begin{itemize}
%  \item Recently, Nguyen/Wu (2013) considered mitxure-of-mixture models based on subcomponents from the multivariate t.
%   \item	label  switching:  Mention  Cron/West, Roodaki?
%  \item ADD Formal ide:
%b)	DiZio (2007), p. 2575 – additional maths and formal ide conditions
%c)	 Rod et al (Sup, 2014), Section 3.3. + 3.4.
%d) Willse et al, ADD litDiZio
%\end{itemize}

\section{Sparse hierarchical mixture of mixtures model}\label{sec:model-specification}

 \subsection{Model definition} \label{sec:Dirprior}

 Following previous work on hierarchical mixtures of mixtures, we
 assume that $N$ observations $\by_i$, $i=1,\ldots,N$ of dimension 
 $\dim(\by_i)=r$  are drawn independently from a finite mixture
 distribution with $K$ components,
\begin{align}
  p(\by_i| \bTheta,\boldeta) &= \sum \limits_{k=1}^{K} \eta_k p_k( \by_i |\btheta_k), \quad \bTheta=(\btheta_1,\ldots,\btheta_K),  \label{eq:mixOfMix1}
\end{align}
with each component distribution  $p_k( \by_i |\btheta_k)$ being  a mixture of $L$ normal subcomponents:
\begin{align}
  p_k(\by_i| \btheta_k) &= \sum \limits_{l=1}^{L} w_{kl} f_{\cN}( \by_i | \bmu_{kl}, \bSigma_{kl}). \label{eq:mixOfMix2}
\end{align}
In order to distinguish the component distributions on the upper level
from the Gaussian components on the lower level, we will refer to the
former ones as ``cluster distributions''.  For clustering the
observations based on Bayes' rule, the cluster weights
$\boldeta=(\eta_1, \ldots, \eta_K)$ and the cluster densities
$p_k( \by_i |\btheta_k)$ on the upper level (\ref{eq:mixOfMix1}) are
relevant.

%In  (\ref{eq:mixOfMix1}), , however,
Since the number of data clusters is unknown and needs to be inferred
from the data, we assume that (\ref{eq:mixOfMix1}) is an overfitting
mixture, i.e.~the {\r specified} number of clusters $K$
  exceeds the number of clusters present in the data. Following the
  concept of sparse finite mixtures
  \citep{Mix:MalsinerWalliFruehwirthGruen2014}, we choose a symmetric
  Dirichlet distribution as prior for the weight distribution,
  i.e.~$\boldeta|e_0 \sim Dir_K(e_0)$, and base our choice of $e_0$ on
  the results of \cite{Mix:RousseauMengerson2011} concerning the
  asymptotic behavior of the posterior distribution of an overfitting
  mixture model. They show that this behavior is determined by the
  hyperparameter $e_0$ of the Dirichlet prior on the weights. In
  particular, they prove that, if $e_0<d/2$, where $d$ is the
  dimension of the {\r cluster-specific parameters} $\btheta_k$,
%(consisting of the subcomponent weights, the subcomponent means and variance-covariance matrices, see Equation~\ref{eq:btheta}),
then the posterior expectation of the weights associated with
superfluous clusters asymptotically converges to zero.

Hence, %to estimate the number of clusters on the cluster level (\ref{eq:mixOfMix1}),
we specify a sparse prior on the cluster weights $\boldeta$ by
choosing $e_0\ll d/2$ so that superfluous clusters are emptied during
MCMC sampling and the number of non-empty clusters on the cluster
level is an estimator for the unknown number of data clusters.  In
this way, the specification of a sparse cluster weight prior in an
overfitting mixture of mixtures model provides an ``automatic tool''
to select the number of clusters, avoiding the expensive computation
of marginal likelihoods as, e.g., in~\citet{fru:est}. %, or other model choice criteria.
 Empirical results in
\cite{Mix:MalsinerWalliFruehwirthGruen2014} indicate that $e_0$ needs
to be chosen very small, e.g.~$e_0=0.001$, to actually empty all
superfluous clusters in the finite sample case.

On the lower level  (\ref{eq:mixOfMix2}),
% $L$ is chosen sufficiently large in order to allow in all clusters for
in each cluster $k$, a semi-parametric approximation of the cluster
distributions is achieved by mixing $L$ multivariate Gaussian
subcomponent densities $f_{\cN}(\by_i | \bmu_{kl}, \bSigma_{kl})$,
$l=1,\ldots,L$, according to the subcomponent weight vector
$\bw_k=(w_{k1},\ldots,w_{kL})$.
% , where  $w_{kl}\geq 0$ and $\sum_{l=1} w_{kl}=1$.
The cluster-specific parameter vector
\begin{align}
  \label{eq:btheta}
  \btheta_k &=(\bw_k,\bmu_{k1},\ldots,\bmu_{kL}, \bSigma_{k1}, \ldots, \bSigma_{kL})
\end{align}
consists of % the weight vector
$\bw_k$ as well as the means $\bmu_{kl}$ and covariance matrices
$\bSigma_{kl}$ of all Gaussian subcomponent densities. $L$ is
typically unknown, but as we are not interested in estimating the
``true'' number of subcomponents $L$ forming the cluster, we only
ensure that $L$ is chosen sufficiently large to obtain an accurate
approximation of the cluster distributions. While
  the choice of $L$ is not crucial to ensure a good model fit as long as
  $L$ is sufficiently large, a too generous choice of $L$
  should be avoided for computational reasons as the computational
  complexity of the estimation increases with the number of
  subcomponents $L$.

By choosing the prior $\bw_k
\sim Dir_L(d_0)$ with $d_0=d/2+2$, the approximation of the cluster
density is obtained by filling all $L$ subcomponents, thus avoiding
empty subcomponents.  This choice is motivated again by the results {\bl of}
\cite{Mix:RousseauMengerson2011} who show that, if $d_0>d/2$, the
posterior density asymptotically handles an overfitting mixture by
splitting ``true'' components into two or more identical components.

\subsection{Identification  through hierarchical priors} \label{priorhier}

When fitting the finite mixture model (\ref{eq:mixOfMix1}) with
semi-parametric cluster densities given by (\ref{eq:mixOfMix2}), we
face a special identifiability problem, since the likelihood is
entirely agnostic about which subcomponents form a cluster. Indeed,
the likelihood is completely ignorant concerning the issue which of
\commentG{the  \KL\  components} belong together, since (\ref{eq:mixOfMix1})
can be written as an expanded Gaussian mixture \commentG{with  \KL\ components}
with weights $\tilde{w}_{kl} = \eta_k w_{kl} $,
\begin{eqnarray}  \label{expmix}
p(\by_i| \bTheta,\boldeta)= %\sum_{k=1}^K  \eta_k p_{k}(\by_i|\thetav_k) =
\sum_{k=1}^K  \sum \limits_{l=1}^{L} \tilde{w}_{kl} f_{\cN}(\by_i  | \bmu_{kl}, \bSigma_{kl}).
\end{eqnarray}
\commentG{These  \KL\  components can be permuted in \KLfac\  different}
ways   and the resulting ordering can be used to group them into $K$
different cluster densities, without changing the mixture likelihood \commentG{(\ref{expmix})}.
Hence, the identification of (\ref{eq:mixOfMix1}), up to label
switching on the upper level, hinges entirely on the prior
distribution.

Subsequently, we suggest a hierarchical prior that addresses these
issues explicitly. Conditional on a set of fixed hyperparameters
$\hyp=(e_0,d_0,c_0,g_0,\bG_0,\bB_0,\bm_0,\bM_0,\nu)$, the weight
distribution $\boldeta|e_0 \sim Dir_K(e_0) $ and the $K$
cluster-specific parameter vectors
$\btheta_k|\hyp \simiid p(\btheta_k|\hyp) $ are independent a~priori,
i.e.:
\begin{eqnarray}
p(\boldeta, \btheta_1,\ldots,\btheta_K| \hyp)=p(\boldeta|e_0) \prod_{k=1}^K p(\btheta_k|\hyp).
\end{eqnarray}
This prior formulation ensures that the $K$ non-Gaussian cluster
distributions of the upper level mixture (\ref{eq:mixOfMix1}) are
invariant to permutations.  Within each cluster $k$, the prior
distribution $p(\btheta_k|\hyp)$ admits the following block
independence structure:
\begin{eqnarray} \label{thetsim}
p(\btheta_k|\hyp) =   p(\bw_k|d_0) p(\bmu_{k1},\ldots,\bmu_{kL}|\bB_0,\bm_0,\bM_0,\nu)
p(\bSigma_{k1}, \ldots, \bSigma_{kL}| c_0,g_0,\bG_0),
\end{eqnarray}
where $\bw_k|d_0 \simiid Dir_L(d_0)$.  Conditional on $\hyp$, the
subcomponent means $\bmu_{k1},\ldots,\bmu_{kL}$ are dependent a~priori
as are the subcomponent covariance matrices
$\bSigma_{k1}, \ldots, \bSigma_{kL}$. However, they are assumed to be
exchangeable to guarantee that within each cluster $k$, the $L$
Gaussian subcomponents in (\ref{eq:mixOfMix2}) can be permuted without
changing the prior.

To create this dependence, a hierarchical \lq\lq random effects\rq\rq\
prior is formulated, where, on the upper level, conditional on the
fixed upper level hyperparameters $(g_0,\bG_0,\bm_0,\bM_0,\nu)$,
cluster specific random hyperparameters ($\bC_{0k}$, $\bb_{0k}$), and
$\bLambda_k=\text{diag}(\lambda_{k1},\ldots,\lambda_{kr})$, are drawn
independently for each $k=1, \ldots, K$ from a set of three
independent base distributions:
\begin{align} \label{hierlev1} \bC_{0k}|g_0,\bG_0 &\simiid
  \cW_r(g_0,\bG_0) , & \bb_{0k}|\bm_0,\bM_0 &\simiid
  \cN_r(\bm_0,\bM_0), & % \\ \nonumber
   (\lambda_{k1}, \ldots, \lambda_{kr})|\nu
  &\simiid \cG(\nu,\nu),
\end{align}
where $\cN_r()$ and $\cW_r()$ denote the $r$-multivariate normal and
Wishart distribution, respectively, and $\cG()$ the gamma
distribution, parametrized such that \comment{$E(\lambda_{kl}|\nu)=1$.}
%as in \cite{Mix:Fruehwirth2006}.

On the lower level, conditional on the cluster specific random
hyperparameters $(\bC_{0k}, \bb_{0k}$, $\bLambda_k)$ and the fixed lower
level hyperparameters $(\bB_{0}, c_{0})$, the $L$ subcomponent means
$\bmu_{kl}$ and covariance matrices $\bSigma_{kl}$ are drawn
independently for all $l=1,\dots,L$:
\begin{eqnarray} \label{priormu}
  \bmu_{kl}|\bB_{0}, \bb_{0k},
  \bLambda_k \simiid \cN_r(\bb_{0k},
  \sqrt{\bLambda_k}\bB_{0}\sqrt{\bLambda_k}), \quad
  \bSigma^{-1}_{kl}|c_{0},\bC_{0k} \simiid \cW_r(c_{0},\bC_{0k}).
\end{eqnarray}

\subsection{Tuning the hyperparameters}\label{sec:priorSubcompCov}

To identify the mixture of mixtures model given in
(\ref{eq:mixOfMix1}) and (\ref{eq:mixOfMix2}) through the prior
defined in Section~\ref{priorhier}, the fixed hyperparameters $\hyp$
have to be chosen carefully.  In addition, we select them in a way to
take the data scaling into account, avoiding the need to standardize
the data prior to data analysis.

First, it is essential to clarify what kind of shapes and forms are
\commentG{aimed at as} cluster {\bl distributions}. We give the following
(vague) characterization of a data cluster: A data cluster is a very
``dense'' region of data points, with possibly no ``gaps'' within the
cluster {\r distribution}, whereas different clusters should be
located well-separated from each other, i.e.~here large ``gaps''
between the cluster {\r distribution}s are desired.  We confine
ourselves to the investigation of clusters with approximately convex
cluster shapes, where the cluster center can be seen as a suitable
representative for the entire cluster. Regarding volume, orientation
or asymmetry of the data clusters we are looking for, no constraints
on the cluster shapes and forms are imposed.

Based on this cluster concept, our aim is to model a dense and
connected cluster distribution by a mixture of normal subcomponents.
{\bl Various} strategies regarding the modeling of the subcomponent means
and covariance matrices could be employed. We decided to allow for
flexible shapes for the single subcomponents, ensuring that they
strongly overlap at the same time. An alternative approach would be to use
constrained simple shaped subcomponents, e.g., subcomponents with
isotropic covariance matrices.  However, in this case a large number
of subcomponents might be needed to cover the whole cluster region and
shrinkage of the subcomponent means toward the common cluster center
may not be possible.  Since then some of the subcomponents have to be
located far away from the cluster center in order to fit also boundary
points, considerable distances have to be allowed between subcomponent
means.  This induces the risk of gaps within the cluster distribution
and a connected cluster distribution may not result. Therefore, in our
approach the cluster distributions are estimated as mixtures of only a
few but unconstrained, highly dispersed and heavily overlapping
subcomponents where the means are strongly pulled toward the cluster
center. In this way, a connected cluster distribution is ensured.

In a Bayesian framework, we need to translate these modeling purposes
into appropriate choices of hyperparameters.  On the upper level, the
covariance matrix $\bM_0$ controls the amount of prior shrinkage of
the cluster centers $\bb_{0k}$ toward the overall data center $\bm_0$,
which we specify as the midpoint of the data. \commentG{To obtain a %uninformative
prior, where the cluster centers $\bb_{0k}$  are}
allowed to be widely spread apart and almost no shrinkage {\bl toward}
$\bm_0$ takes place, we choose $\bM_0 \gg \bS_y$, where $\bS_y$ is the
sample covariance matrix of all data, e.g.~$\bM_0 = 10
\bS_y$.

\begin{figure}[t!]
 \centering
    \includegraphics[width=0.3\textwidth, trim = 0 20 0 55, clip]{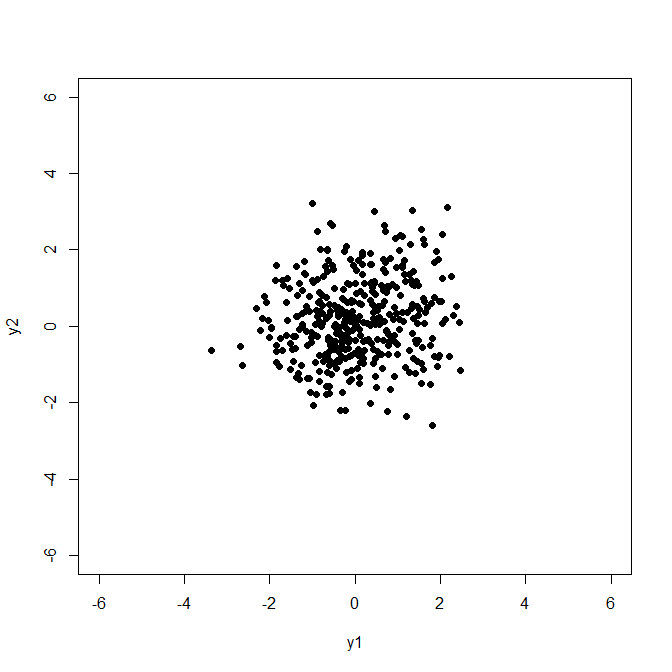}
    \includegraphics[width=0.3\textwidth, trim = 0 20 0 55, clip]{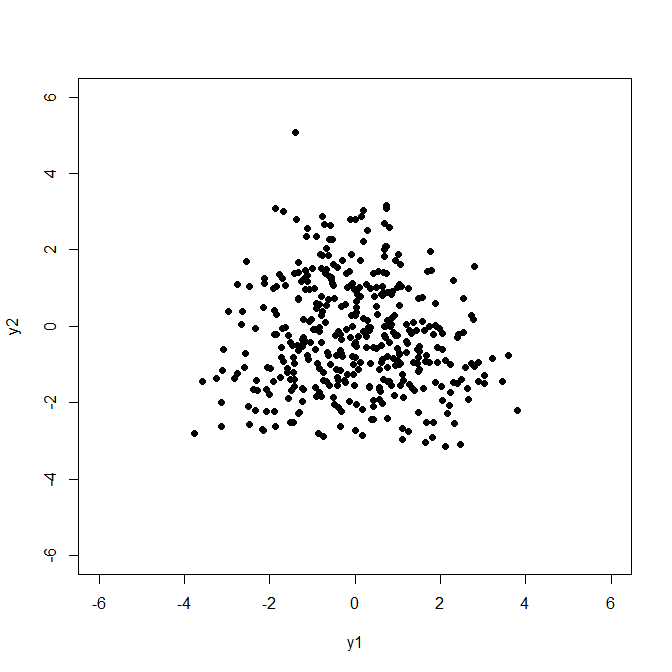}
    \includegraphics[width=0.3\textwidth, trim = 0 20 0 55, clip]{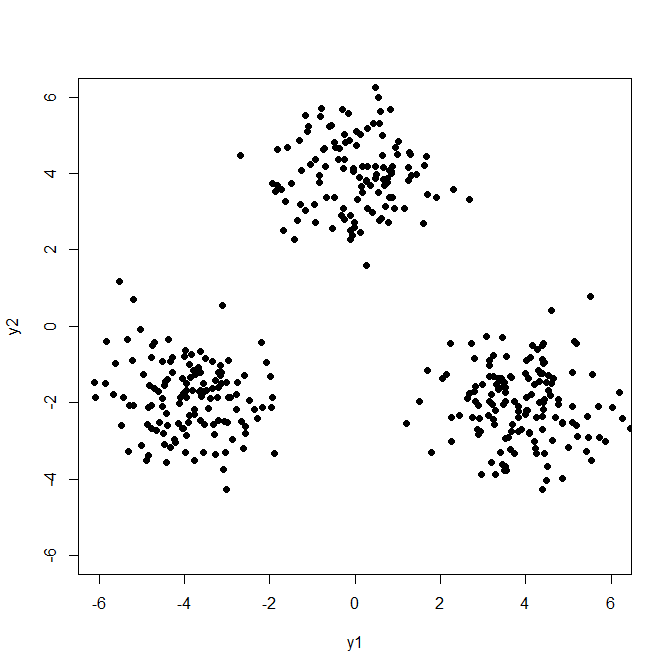}
  \caption{\footnotesize Variance decomposition of a
    mixture distribution. Scatter plots of samples from a
    standard normal mixture distribution with three components and
    equal weights, with a varying amount of heterogeneity $\phi_B$
    explained by the variation of the component means, $\phi_B=0.1$,
    $\phi_B=0.5$ and $\phi_B=0.9$ (from left to
    right).}\label{plot:VarDecom}
\end{figure}

% The hyperparameters
% $c_0, g_0 $,  and $\bG_0$ of the prior on the subcomponent covariance matrix
% $\bSigma_{kl}$ influence the volume and the flexibility of the
% subcomponent densities.  The scale matrix $\bB_{0}$ of
% the prior on the subcomponent means $\bmu_{kl}$ controls how dispersed
% from the cluster center $\bb_{0k}$ the single subcomponent means are
% allowed to be. Pulling the subcomponent means to the cluster center is
% important because the more the subcomponent means are allowed to
% differ from $\bb_{0k}$, the higher is the risk of gaps within a
% cluster. On the other hand, the diagonal matrix $\bLambda_k$ should act as a local adjustment factor
% which allows to correct the covariance matrix $\bB_0$ for each cluster and each dimension at a small scale.
Our strategy for appropriately specifying the hyperparameters $\bG_0$ and  $\bB_{0}$
% the hyperparameters for the lower level
is based on the variance decomposition of the mixture of mixtures
model, which splits $\mathit{Cov}(\bY)$ into the different sources of
variation. For a finite mixture model with $K$ clusters, as given
in~\eqref{eq:mixOfMix1}, the total heterogeneity $\mathit{Cov}(\bY)$
can be decomposed in the following way
\citep[p.~170]{Mix:Fruehwirth2006}:
\begin{equation} \label{vardec} \mathit{Cov}(\bY) = \sum_{k=1}^K
\eta_k \bSigma_{k} + \sum_{k=1}^K \eta_k \bmu_k \bmu_k'- \bmu
\bmu'=(1-\phi_B)\mathit{Cov}(\bY) + \phi_B \mathit{Cov}(\bY),
\end{equation} where the cluster means $\bmu_k$ and the cluster
covariance matrices $\bSigma_k$ are the first and second moments of
the cluster distribution $p_k( \by_i |\btheta_k)$ and $\bmu=\sum_k
\eta_k \bmu_k$ is the mixture mean. In this decomposition $\phi_B$ is
the proportion of the total heterogeneity explained by the variability
of the cluster means $\bmu_k$ and $(1-\phi_B)$ is the proportion
explained by the average variability within the clusters. The larger
$\phi_B$, the more the clusters are separated, {\bl as  illustrated} in
Figure~\ref{plot:VarDecom} for a three-component standard Gaussian
mixture with varying values of $\phi_B$.

For a mixture of mixtures model, the heterogeneity
$(1-\phi_B)\mathit{Cov}(\bY)$ explained within a cluster can be split
further into two sources of variability, namely the proportion
$\phi_W$ explained by the variability of the subcomponent means
$\bmu_{kl}$ around the cluster center $\bmu_k$, and the proportion
$(1-\phi_W)$ explained by the average variability within the
subcomponents:
% If we further assume for simplicity that the within-cluster
% variability is the same for all clusters,
% i.e.~$\bSigma_1=\ldots=\bSigma_K$, and also that the variability
% explained by the different subcomponents is the same,
% i.e.~$\bSigma_{kl}\equiv\bSigma_{11}$, for all $k$ and $l$, the
% following decomposition of the variability in a mixture of mixtures
% model is obtained:
\begin{align}   \nonumber
  \mathit{Cov}(\bY) & =\sum_{k=1}^K \eta_k \bSigma_{k} + \sum_{k=1}^K \eta_k \bmu_k \bmu_k'- \bmu \bmu' \\
  &=   \sum_{k=1}^K \eta_k\sum_{l=1}^L w_{kl}\bSigma_{kl}  +
  \sum_{k=1}^K \eta_k\left(\sum_{l=1}^L w_{kl} \bmu_{kl}\bmu_{kl}' - {\bmu}_k{\bmu}_k'\right) +  \sum_{k=1}^K \eta_k \bmu_k \bmu_k'- \bmu \bmu'  \label{vardecmixmix} \\
  &= (1-\phi_W)(1-\phi_B)\mathit{Cov}(\bY) + \phi_W (1-\phi_B)\mathit{Cov}(\bY)  + \phi_B \mathit{Cov}(\bY). \nonumber
\end{align}
Based on this variance decomposition we select the
proportions $\phi_B$ and $\phi_W$ and incorporate them into the
specification of the hyperparameters of our hierarchical prior.

$\phi_B$ defines the proportion of variability explained by the
different cluster means.  We suggest to specify $\phi_B$ not too
large, e.g., to use $\phi_B=0.5$. This specification may seem to be
counterintuitive as in order to model well-separated clusters it would
seem appropriate to select $\phi_B$ large. However, if $\phi_B$ is
large, the major part of the total heterogeneity of the data is
already explained by the variation (and separation) of the cluster
means, and, as a consequence, only a small amount of heterogeneity is
left for the within-cluster variability. This within-cluster
variability in turn will get even more diminished by the variability
explained by the subcomponent means leading to a small amount of
variability left for the subcomponents. Thus for large values  of
$\phi_B$, estimation of tight subcomponent densities would result,
undermining our modeling aims.

$\phi_W$ defines the proportion of within-cluster variability
explained by the subcomponent means. $\phi_W$ also controls how strongly the
subcomponent means are pulled together and influences the overlap of the subcomponent densities. To
achieve strong shrinkage of the subcomponent means toward the cluster
center, we select small values of $\phi_W$, e.g.~$\phi_W=0.1$. Larger
values of $\phi_W$ may introduce gaps within a cluster, which we want
to avoid.

Given $\phi_B$ and $\phi_W$, we specify the scale matrix $\bG_{0}$ of
the prior on $\bC_{0k}$ such that the a~priori expectation of the first
term in the variance decomposition (\ref{vardecmixmix}), given by
\begin{eqnarray*}
  E\left( \sum_{k=1}^K \eta_k\sum_{l=1}^L w_{kl}\bSigma_{kl} \right)=
 \sum_{k=1}^K E( \eta_k) \sum_{l=1}^L E(w_{kl}) E(E(\bSigma_{kl}|\bC_{0k}))= g_0/ (c_0-(r+1)/2)  \bG_0^{-1},
\end{eqnarray*}
  matches the desired
amount of heterogeneity explained by a subcomponent:
\begin{eqnarray}  \label{priorG0}
  g_0/(c_0-(r+1)/2)  \bG_0^{-1} =(1-\phi_W)(1-\phi_B)\mathit{Cov}(\bY).
\end{eqnarray}
{\r We replace $\mathit{Cov}(\bY)$ in (\ref{priorG0}) with the main diagonal of the sample
  covariance $\bS_y$ to take only the scaling of the data into account
  \citep[see~e.g.~][]{Mix:Fruehwirth2006}. % \citep{Mix:Stephens1997,Mix:Fruehwirth2006},
This gives} the following specification for $\bG_0$:
\begin{equation}\label{eq:diagG0}
  \bG_0^{-1}= (1-\phi_W)(1-\phi_B) (c_0-(r+1)/2)/g_0 \cdot \text{diag}( \bS_y ).
\end{equation}
Specification of the prior of the subcomponent covariance matrices
$\bSigma_{k1}, \ldots, \bSigma_{kL}$ is completed by defining the
scalar prior hyperparameters $c_{0}$ and $g_0$.  \citet[Section 6.3.2,
p.~192]{Mix:Fruehwirth2006} suggests to set $c_{0}> 2 +(r-1)/2$. In
this way the eigenvalues of $\bSigma_{kl}\bSigma_{km}^{-1}$ are
bounded away from $0$ avoiding singular matrices. We set
$c_{0}= 2.5 +(r-1)/2$ to allow for a large variability of
$\bSigma_{kl}$. {\r }The Wishart density is regular if
$g_0>(r-1)/2$ {\r and} in the following we set $g_0=0.5+(r-1)/2$.

Regarding the prior specification of the subcomponent means
$\bmu_{k1}, \ldots, \bmu_{kL}$, we select the scale matrix $\bB_{0}$ in order to
concentrate a lot of mass near the cluster center $\bb_{0k}$, pulling
$\bmu_{kl}$  {\bl toward} $\bb_{0k}$.  Matching the a~priori
expectation of the second term in the variance decomposition
(\ref{vardecmixmix}), given by
\begin{eqnarray*}
  E \left( \sum_{k=1}^K  \eta_k \left(\sum_{l=1}^L w_{kl} \bmu_{kl}\bmu_{kl}' - {\bmu}_k{\bmu}_k' \right) \right)=
 \sum_{k=1}^K E( \eta_k) \sum_{l=1}^L E(w_{kl}) E(\bmu_{kl}\bmu_{kl}' - {\bmu}_k{\bmu}_k')= \bB_{0},
\end{eqnarray*}
  to the desired
proportion of explained heterogeneity  and, using once more only the main diagonal
of $\bS_y$ we obtain $\bB_{0}= \phi_W (1-\phi_B) \text{diag}(\bS_y)$,
%\begin{align*}
%  \bB_{0}&= \phi_W (1-\phi_B) \text{diag}(\bS_y),
%\end{align*}
 which incorporates our idea that only a small proportion $\phi_W$ of
the within-cluster variability should be explained by the variability
of the subcomponent means.

After having chosen $\phi_B$ and $\phi_W$, basically the cluster
structure and shape is a~priori determined. However, in order to allow
for  %slightly
more flexibility in capturing the unknown cluster shapes
in the sense that within each cluster the amount of shrinkage of the
subcomponent means $\bmu_{kl}$ toward the cluster center $\bb_{0k}$
{\bl need} not to be the same for all dimensions, for each cluster $k$ and
each dimension $j$ additionally a random adaptation factor
$\lambda_{kj}$ is introduced in (\ref{priormu}) which adjusts
$\bB_{0}$.  The gamma prior for $\lambda_{kj}$ in (\ref{hierlev1})
implies that the prior expectation of the covariance matrix of
$\bmu_{kl}$ equals $\bB_{0}$. However, $\lambda_{kj}$ acts as a local
adjustment factor for cluster $k$ which allows to shrink (or inflate)
the variance of subcomponent means $\mu_{klj}$ in dimension $j$ in
order to adapt to a more (or less) dense cluster distribution as
specified by $\bB_{0}$. In order to allow only for small adjustments
of the specified $\bB_{0}$, we choose $\nu=10$, in this way almost
{\r 90\%} of the a~priori values of $\lambda_{kj}$ are between $0.5$ and
% > diff(pgamma(c(0.5, 1.5), 10, 10))
% [1] 0.8983183
$1.5$. \commentG{This} %It should be noted that t
 hierarchical prior specification
for $\bmu_{kl}$ corresponds to the normal gamma prior
\citep{Mix:GriffinBrown2010} which has been applied
 by \cite{Mix:Fruehwirth2011} and
\cite{Mix:MalsinerWalliFruehwirthGruen2014} in the context of finite
mixture models for variable selection.

\subsection{Relation to BNP mixtures} \label{BNP}

Our approach bears resemblance to various approaches in BNP modeling.
First of all, the concept of sparse finite mixture{\r s} as used in
\citet{Mix:MalsinerWalliFruehwirthGruen2014} is related to Dirichlet
process (DP) mixtures \citep{Mix:Mueller+Mitra:2013} where the discrete mixing
distribution in the finite mixture (\ref{eq:mixOfMix1}) is substituted
by a random distribution $G \sim DP(\alpha, H) $, drawn from a DP
prior with precision parameter $\alpha$ and base measure $H$. As a
draw $G$ from a DP is almost surely discrete, the corresponding model
has a representation as an infinite mixture:
\begin{align}
  p(\by) %| \bTheta,\boldeta)
  &= \sum \limits_{k=1}^{\infty} \eta_k p_k( \by |\btheta_k), \label{eq:infMix1}
  %\quad \bTheta=(\btheta_1,\ldots,\btheta_K),
\end{align}
with \iid\ atoms $\btheta_k \simiid H $ drawn from the base measure
$H$ and \commentG{weights $ \eta_k=\stick_k \prod_{j=1}^{k-1} (1- \stick_j )$ %, k=1, 2,\ldots$,
obeying the stick breaking representation
with $\stick_k \simiid \Betadis{1,\alpha}$ \citep{set:con}.}

If the hyperparameter in the weight distribution $\boldeta$ of a
sparse finite mixture is chosen as $e_0=\alpha/K$, i.e.~$\boldeta
\sim Dir_K(\alpha/K)$, and the component parameters $\btheta_k \simiid
H $ are \iid\ draws from $H$, then as $K$ increases, the sparse finite
mixture in Equation~(\ref{eq:mixOfMix1}) converges to a DP mixture
with mixing distribution $G \sim DP(\alpha,H)$, see
\citet{gre-ric:mod}.  For example, the sparse finite Gaussian mixture
introduced in \citet{Mix:MalsinerWalliFruehwirthGruen2014} converges
to a Dirichlet process Gaussian mixture as $K$ increases, with
$(\bmu_{k},\bSigma_{k})$ being \iid\ draws from the appropriate base
measure $H$.

The more general sparse finite mixture of mixtures model introduced in
this paper also converges to a Dirichlet process mixture where the
atoms are finite mixtures indexed by the parameter $\btheta_k$ defined
in (\ref{eq:btheta}).  The parameters $\btheta_k$ are \iid\ draws from
the base measure (\ref{thetsim}), with strong dependence among the
means $\bmu_{k1},\ldots,\bmu_{kL}$ and covariances
$\bSigma_{k1}, \ldots, \bSigma_{kL}$ within each cluster $k$.  This
dependence is achieved through the two-layer hierarchical prior
described in (\ref{hierlev1}) and (\ref{priormu}) and is essential to
create well-connected clusters from the subcomponents, as outlined in
Section~\ref{sec:priorSubcompCov}.

Also in the BNP framework models have been introduced that create
dependence, either in the atoms and/or in the weights attached to the
atoms.
% \citet{teh-etal:hie} introduced the hierarchical DP prior over a
% base measure which has more general atoms than in a standard DP
% mixture, namely each atom is itself a Dirichlet process
% $DP(\alpha,H)$. A single draw $H \sim DP(\alpha,H)$ is used to
% define the base measure of another Dirichlet process
% $DP(\beta,H)$. The data form known groups, $y_{it}, t=\ldots, N_i$
% with clusters in each group.
For instance, the nested DP process of
\citet{Mix:RodriguezDunsonGelfand2008} allows to cluster distributions
across $N$ units. Within each unit $i$, $i=1,\ldots,N$, repeated
(univariate) measurements $y_{it}, t=1,\ldots,
N_i$ %$y_{it} \simiid F_i $
arise as independent realizations of
a %non-parametric distribution $F_i$, being a
DP Gaussian mixture with random mixing distribution $G_i$.  The $G_i$s
are \iid\ draws from a DP, in which the base measure is itself a
Dirichlet process $DP(\beta,H)$, i.e.~$G_i \simiid DP(\alpha,
DP(\beta,H))$. Hence, two distributions $G_i$ and $G_j$ either share
the same weights and atoms sampled from $H$, or the weights and atoms
are entirely different.  If only a single observation $y_i$ is
available in each unit, i.e.~$N_i=1$, then the nested DP is related to
our model. In particular, it has a two-layer representation as in
(\ref{eq:mixOfMix1}) and (\ref{eq:mixOfMix2}), however with both $K$
and $L$ being infinite.  The nested DP can, in
principal, be extended to multivariate observations $\by_i$. In this case,  $p(\by_i)$ takes the same form as in
(\ref{eq:infMix1}), with the same stick breaking representation for
the cluster weights $\eta_1,\eta_2,\ldots$. % and $DP(\alpha,H)$
On the lower level, each cluster distribution $p_k( \by_i |\btheta_k)$
is a DP Gaussian mixture:
\begin{align}
  p_k(\by_i| \btheta_k) &= \sum \limits_{l=1}^{\infty} w_{kl} f_{\cN}( \by_i | \bmu_{kl}, \bSigma_{kl}), \label{eq:infMix2}
\end{align}
where the component weights $ w_{kl}$ are derived from the stick
breaking representation
$ w_{kl}=\stickw_{kl} \prod_{j=1}^{l-1} (1- \stickw_{kj} ), l=1,
2,\ldots$
where $\stickw_{kl} \simiid \Betadis{1,\beta}$. For the nested DP,
dependence is introduced only on the level of the weights and sticks,
as the component parameters $\bmu_{kl}, \bSigma_{kl} \simiid H $ are
\iid\ draws from the base measure $H$.   This  lack
of prior dependence among the atoms $(\bmu_{kl}, \bSigma_{kl})$ is
likely to be an obstacle in a clustering context.

The BNP approach most closely related to our model is the infinite
mixture of infinite Gaussian mixtures (I$^2$GMM) model of
\citet{Mix:Yerebakan2014} which also deals with clustering
multivariate observations from non-Gaussian component
densities.\footnote{We would like to thank a reviewer for pointing us
  to this paper.}
The I$^2$GMM model has a two-layer hierarchical representation like
the nested DP. On the top level, \iid\ cluster specific locations
$\bb_{0k}$ and covariances $\bSigma_{k}$ are drawn from a random
distribution $G \sim DP(\alpha,H)$ arising from a DP prior with base
measure $H$ being equal to the conjugate normal-inverse-Wishart
distribution.  A cluster specific DP is introduced on the lower level
as for the nested DP; however, the I$^2$GMM model is more flexible, as
prior dependence is also introduced among the atoms belonging to the
same cluster.  More precisely, $\ym_i \sim \cN_r(\bmu _i,\bSigma_k)$,
with $\bmu _i \simiid G_k$, where $G_k \sim DP(\beta,H_k)$ is a draw
from a DP with cluster specific base measure
$H_k = \cN_r (\bb_{0k},\bSigma_k/\kappa_1)$.

It is easy to show that  the I$^2$GMM model has an infinite
two-layer representation as in (\ref{eq:infMix1}) and
(\ref{eq:infMix2}), with exactly the same stick breaking
representation.\footnote{Note that the notation in
  \citet{Mix:Yerebakan2014} is slightly different, with $\gamma$ and
  $\alpha$ corresponding to $\alpha$ and $\beta$ introduced above.}
However, the I$^2$GMM model has a constrained form on the lower level,
with homoscedastic covariances $\bSigma_{kl} \equiv \bSigma_{k}$,
whereas the locations $\bmu_{kl}$ scatter around the \commentG{cluster centers}  $\bb_{0k}$ as in our model:
   \begin{eqnarray}  \label{priorIMM}
   (\bb_{0k}, \bSigma_{k}) \simiid H,  \qquad
   \bmu_{kl}| \bb_{0k},\bSigma_k  \simiid H_k.
 \end{eqnarray}
 In our sparse mixture of mixtures model, we found it useful to base
 the density estimator on heteroscedastic covariances $\bSigma_{kl}$,
 to better accommodate \commentG{the} non-Gaussianity of the cluster densities with a
 fairly small number $L$ of subcomponents.  It should be noted that
 our semi-parametric density estimator is allowed to display
 non-convex shapes, as illustrated in Figure~C.2 in the Appendix.
 Nevertheless, we could have considered a mixture in
 (\ref{eq:mixOfMix2}) where $\bSigma_{kl} \equiv \bSigma_{k}$, with
 the same base measure for the atoms
 $(\bmu_{k1},\ldots,\bmu_{kL},\bSigma_{k})$ as in
 (\ref{priorIMM}). %, or a mixture where atoms $(\bmu_{k1},\bSigma_{kl})$ are \iid\ draws from the same base measure as in \citet{Mix:RodriguezDunsonGelfand2008}.
 In this case, the relationship between our sparse finite mixture and
 the I$^2$GMM model would become even more apparent: by choosing
 $e_0=\alpha/K$ and $d_0=\beta/L$ and letting $K$ and $L$ go to
 infinity, our model would converge to the I$^2$GMM model.

\section{Clustering and posterior inference}\label{sec:modelEstimation}

\subsection{Clustering and selecting the number of clusters}\label{sec:clust}

For posterior inference, two sequences of allocation variables are introduced, namely the cluster assignment indicators $\bS=(S_1,\ldots,S_N)$ and the within-cluster allocation variables
$\bI=(I_1, \ldots, I_N)$.
% More specifically, for each MCMC iteration $m=1,\ldots,M$, the
% corresponding posterior draw $S_i ^{(m)} \in \{1,\ldots,K\} $
% assigns each observation $\by_i$ to cluster $S_i ^{(m)}$ on the
% upper level of the mixture of mixtures model. On the lower level,
% $I_i ^{(m)} \in \{1,\ldots,L\} $ assigns all observations $\by_i$
% within a specific cluster $k$ (i.e.~$S_i=k$) to subcomponent
% $I_i ^{(m)}$. Hence, the pair $(S_i ^{(m)}, I_i ^{(m)})$ carries all
% the information needed toassign each observation to a unique
% component in the expanded mixture (\ref{expmix}).
More specifically, $S_i \in \{1,\ldots,K\} $ assigns each observation
$\by_i$ to cluster $S_i$ on the upper level of the mixture of mixtures
model. On the lower level, $I_i \in \{1,\ldots,L\} $ assigns
{\r observation $\by_i$ to subcomponent $I_i $}. Hence, the pair $(S_i ,
I_i )$ carries all the information needed to assign each observation
to a unique component in the expanded mixture (\ref{expmix}).

Note that for all observations $\ym_i$ and $\ym_j$ belonging to the
same cluster, the \commentG{upper level indicators
  $S_i=S_{j}$} %$S_i=S_{j}=k$
will be the same, while the lower level indicators $I_i\neq I_{j}$
might be different, meaning that they belong to different
subcomponents within the \comment{same} cluster. It should be noted
that the Dirichlet prior $\bw_k \sim Dir_L(d_0)$, with $d_0>d/2$, on
the weight distribution ensures overlapping densities within each
cluster, in particular if $L$ is overfitting. Hence the indicators
\comment{$I_i$} %$I_i|S_i=k$
will typically cover all possible values $\{1,\ldots,L\}$
\comment{within each cluster}.

For clustering, only the upper level indicators $\bS$ \commentG{are  %will be
explored, %automatically
integrating implicitly} over the uncertainty of assignment
to the subcomponents on the lower level.  A cluster
$\Cl_k =\{i| S_i = k\}$ is thus a subset of the data indices
$\{1,\ldots,N\}$, containing all observations with identical
\commentG{upper level indicators}. Hence, the indicators $\bS$ define a random partition
$\parti=\{\Cl_1,\ldots, \Cl_{\Kn}\}$ of the $N$ data points in the
sense of \citet{lau-gre:bay}, as $\by_i$ and $\by_{j}$ belong to the
same cluster, {\bl if and only if} $S_i=S_{j}$.  The partition $\parti$ contains
$\Kn=|\parti|$ clusters, where $|\parti|$ is the cardinality of
$\parti$.  Due to the Dirichlet prior $\boldeta \sim Dir_K(e_0)$, with
$e_0$ close to 0 {\r to obtain a sparse finite mixture}, $K_0$ is a random number
being a~priori much smaller than $K$.

For a sparse finite mixture model with $K$ clusters, the prior
distribution over all random partitions $\parti$ of $N$ observations
is derived from the joint (marginal) prior
%\begin{eqnarray} \label{margS}
 $ p(\Siv) = \int \prod_{i=1}^N p(S_i| \etav) d \, \etav$ \commentG{which
%\end{eqnarray}
 is given, e.g., in} \citet[p.~66]{Mix:Fruehwirth2006}:
\begin{eqnarray}
\displaystyle p( \Siv)
% = \frac{\Gamfun{ K \ed{0}} \prod_{k=1}^K \Gamfun{N_k(\Siv)+\ed{0}}}{ \Gamfun{N+K \ed{0}} \Gamfun{\ed{0}}^K} %= \frac{\Gamfun{K \ed{0}}}{\Gamfun{N+K \ed{0}}\Gamfun{\ed{0}}^{\Kn}} \prod_{k: N_k(\Siv)>0} \Gamfun{N_k(\Siv)+\ed{0}} ,
= \frac{\Gamfun{K \ed{0}}}{\Gamfun{N+K \ed{0}}\Gamfun{\ed{0}}^{\Kn}} \prod_{k: N_k>0} \Gamfun{N_k+\ed{0}} ,
\label{intro:dirich:ml}
\end{eqnarray}
% $N_k(\Siv)= \Count{S_i=k}$.
where $N_k = \Count{S_i=k}$.
For a given partition $\parti$ with $\Kn$ data clusters, there are $K!/(K-\Kn) !$ assignment vectors
$\Siv$ that belong to the equivalence class defined by $\parti$. The prior {\bl distribution} over all random partitions $\parti$ is then obtained by summing over all assignment vectors $\Siv$ that belong to the equivalence class defined by $\parti$:
\begin{eqnarray}
\displaystyle p(\parti| \Kn) = \frac{K!}{(K-\Kn) !}
 \frac{\Gamfun{K \ed{0}} }{\Gamfun{N+ K \ed{0}}
 \Gamfun{\ed{0}}^{\Kn}} \prod_{k: N_k>0 } \Gamfun{N_k+\ed{0}}, \label{ijhuh:ml}
\end{eqnarray}
which takes the form of a product partition model and therefore is invariant to permuting the cluster labels.
Hence, it is possible to derive the prior predictive distribution
$p(S_i|\Siv_{-i})$, where $\Siv_{-i}$ denote all indicators, excluding $S_i$.  Let $\Kni$ be the number of non-empty clusters implied by $\Siv_{-i}$ and let $\Nki{k}$ be the corresponding cluster sizes.
From (\ref{intro:dirich:ml}), we  obtain the following  probability
 that $S_i$ is assigned to an existing cluster $k$:%, with $\Nki{k}>0$:
\begin{eqnarray}
\displaystyle \Prob{S_i=k|\Siv_{-i}, \Nki{k} >0} % = \frac{p(\Siv)}{p(\Siv_{-i})}
 = \frac{\Nki{k} + \ed{0}}{N-1 + \ed{0} K}. \label{seq}
\end{eqnarray}
%whereas the probability that observation $S_i$ is assigned to an empty cluster $k$ with $k \in I=\{k: N_k(\Siv_{-i})=0 \}$ is equal to
%\begin{eqnarray} \displaystyle \Prob{S_i =k|\Siv_{-i}, k  \in I} =   \frac{p(\Siv)}{p(\Siv_{-i})} = \frac{\ed{0}}{N-1 + \ed{0} K}.  \label{seq} \end{eqnarray}
 The prior probability that  $S_i$ creates a new cluster with $S_i \in I =\{k| \Nki{k}=0 \} $ is equal to
\begin{eqnarray}
\displaystyle \Prob{S_i \in I |\Siv_{-i}} =  (K-\Kni) \Prob{S_i =k^*|\Siv_{-i}, k^*  \in I} =
\frac{\ed{0} (K-\Kni)}{N-1 + \ed{0} K}.  \label{seq_empty}
\end{eqnarray}
% since there are $K-\Kni$ empty clusters.
%
%Furthermore, the number $K_0$ of non-empty clusters can be easily identified as the number of unique values $k_1,\ldots, k_{K_0}$ among the $N$ indicators $\bS$.
It is illuminating to investigate the prior probability to create new
\commentG{clusters %within a sparse finite mixture model
in detail}.  First of
all, for $\ed{0}$ independent of $K$, this probability not only
depends on $\ed{0}$, but also increases with $K$. Hence a sparse
finite mixture model based on the prior
$\etav \sim \Dirinv{K}{\ed{0}}$ can be regarded as a two-parameter
model, where both $\ed{0}$ and $K$ influence the \commentG{a~priori expected} number of data clusters $\Kn$ which is determined for a DP mixture
solely by $\alphaDP$.  A BNP two-parameter mixture is obtained from
the Pitman-Yor process (PYP) prior $PY{(\betaPY,\alphaPY)}$ with
$\betaPY \in [0,1), \alphaDP > -\betaPY $ \citep{pit-yor:two}, with
stickbreaking representation
$\stick_k \simiid \Betadis{1-\betaPY,\alphaDP+k\betaPY}$. The DP prior
results as that special case where $\betaPY=0$.

Second, the prior probability (\ref{seq_empty}) to create new clusters
in a sparse finite mixture model decreases, as the number $\Kni$ of
non-empty clusters increases. This is in sharp contrast to DP mixtures
where this probability is constant and PYP mixtures where this
probability increases, see e.g., \citet{fal-bar:gib}.

Finally, what distinguishes a sparse finite mixture model, both from a
DP as well as a PYP mixture, is the a~priori expected number of data
clusters $\Kn$, as the number $N$ of observations increases. For $K$
and $\ed{0}$ independent of $N$, the probability to create new
clusters decreases, as $N$ increases, and converges to 0, as $N$ goes
to infinity.  Therefore, $\Kn$ is asymptotically independent of $N$
for sparse finite mixtures, whereas for the DP process
$\Kn \sim \alphaDP \log (N)$ \citep{Mix:Korwar+Hallnder:1973} and
$\Kn \sim N ^ \betaPY$ obeys a power law for PYP mixtures
\citep{fal-bar:gib}.  This leads to quite different clustering
behavior for these three types of mixtures.

A well-known limitation of DP priors is that a~priori the cluster
sizes are expected to be geometrically ordered, with one big cluster,
geometrically smaller clusters, and many singleton clusters
\citep{Mix:Mueller+Mitra:2013}.  PYP mixtures are known to be more
useful than the DP mixture for data with many significant, but small
\commentG{clusters}. %weights.
 A common criticism concerning finite mixtures is that the
number of clusters needs to be known a~priori. Since this is not the
case for sparse finite mixtures, they are useful in the context of
clustering, in particular in cases where the data arise from a
moderate number of clusters, that does not increase as the number of
data points $N$ increases.

\subsection{MCMC estimation and posterior inference}

Bayesian estimation of the sparse hierarchical mixture of mixtures
model is performed using MCMC methods based on data augmentation and
Gibbs sampling. We only need standard Gibbs sampling steps, see the
detailed MCMC sampling scheme in Appendix~A.

In order to perform inference based on the MCMC draws, i.e.~to
cluster the data, to estimate the number of clusters, to solve the
label switching problem on the higher level and to estimate
cluster-specific parameters, several existing procedures can be easily
adapted and applied to post-process the posterior draws of a mixture
of mixtures model, e.g., those which are, for instance, implemented in
the \proglang{R} packages \pkg{PReMiuM} \citep{Mix:Liverani+Hastie+Azizi:2015}
and \pkg{label.switching} \citep{Mix:Papastamoulis:2015}.

For instance, the approach in \pkg{PReMiuM} is based on the posterior
probabilities of co-clustering, expressed through the similarity
matrix $\Prob{S_i=S_j|\ym}$ which can be estimated from the \commentG{$M$} posterior
 draws $\bS ^{(m)}, m=1,\ldots,M $, see Appendix~B for details. The
methods implemented in \pkg{label.switching} aim at resolving the
label switching problem when fitting a finite mixture model using
Bayesian estimation. Note that in the case of the mixture of mixtures
model label switching occurs on two levels. On the cluster level, the
label switching problem is caused by invariance of the mixture
likelihood given in Equation (\ref{eq:mixOfMix1}) with respect to
reordering of the clusters.  On this level, label switching has to be
resolved, since the single cluster distributions need to be
identified.  On the subcomponent level, label switching happens due to
the invariance of Equation (\ref{eq:mixOfMix2}) with respect to
reordering of the subcomponents.  As we are only interested in
estimating the entire cluster distributions, it is not necessary to
identify the single subcomponents. Therefore, the label switching
problem can be ignored on this level.

In this paper, the post-processing approach employed first performs a
model selection step. The posterior draws of the indicators
$\bS ^{(m)}, m=1,\ldots,M $ are used to infer the number of non-empty
clusters $\Kn ^{(m)}$ on the upper level of the mixture of mixtures model
and the number of data clusters is then estimated as the mode.
Conditional on the selected model, an identified model is obtained
based on the point process representation of the estimated
mixture. This method was introduced in
\citet[p.~96]{Mix:Fruehwirth2006} and successfully applied to
model-based clustering in various applied research, see
e.g.~\cite{fru:pan} for some review.  This procedure has been adapted
to sparse finite mixtures in \cite{Mix:Fruehwirth2011} and
\cite{Mix:MalsinerWalliFruehwirthGruen2014} and is easily extended to
deal with sparse mixture of mixtures models, see Appendix~B for more
details. We will use this post-processing approach in our simulation
studies and the applications in Section~\ref{sec:applications} and
Appendices~C, D and F to determine a partition of the data based on the
maximum a~posteriori (MAP) estimates of the relabeled cluster
assignments.

\section{Simulation studies and applications}\label{sec:applications}
The performance of the proposed strategy for
selecting the unknown number of clusters and identifying the cluster
distributions is illustrated in two simulation studies.  In the first simulation study we
investigate whether we are able to capture dense non-Gaussian data
clusters and estimate the true number of data clusters. Furthermore,
the influence of the specified maximum number of clusters $K$ and
subcomponents $L$ on the clustering results is studied. In the second
simulation study the sensitivity of the a~priori defined proportions
$\phi_B$ and $\phi_W$ on the clustering result is investigated.  For a
detailed description of the simulation design and results see
Appendix~C. Overall, the results indicated that our approach performed
well and yielded promising results.

To further evaluate our approach, we fit the sparse hierarchical
mixture of mixtures model on benchmark data sets and real data.  First,
we consider five data sets which were previously used to benchmark
algorithms in cluster analysis.  For these data sets we additionally
apply the ``merging strategy'' proposed by
\cite{Mix:BaudryRafteryCeleuxLoGottardo2010} in order to compare the
results to those of our approach. For these benchmark data sets class
labels are available and we assess the performance by comparing how
well our approach is able to predict the class labels using the
cluster assignments, measured by the misclassification rate as well as
the adjusted Rand index.

To assess how the algorithm scales to larger data sets we investigate
the application to two flow cytometry data sets. The three-dimensional
DLBCL data set \citep{Mix:LeeMcLachlan2013b} consists of around 8000
observations and comes with manual class labels which can be used as
{\r benchmark}.  The GvHD data set \citep{Mix:Brinkman2007} consists
of 12441 observations, but no class labels are
available. \commentG{We compare the clusters detected
  for this data set qualitatively to solutions previously reported in
  the literature.}

The detailed description of all investigated data sets as well as of
the derivation of the performance measures are given in Appendix~D.
For the benchmark data sets, the number of estimated clusters
$\hat{K}_0$, the adjusted Rand index ($\mathit{adj}$), and
misclassification rate ($\mathit{er}$) are reported in
Table~\ref{table:benchmarkNEW} for all estimated models.  In the first
columns of Table~\ref{table:benchmarkNEW},  the name of the data set,
the number of observations $N$, the number of variables $r$ and the
number of \commentG{true} classes $K^{true}$ (if known) are reported.
To compare our approach to the merging approach proposed by
\cite{Mix:BaudryRafteryCeleuxLoGottardo2010}, we use the function
\texttt{Mclust} of the \proglang{R} package \pkg{mclust}
\citep{Mix:FraleyRafteryScrucca2012} to first fit a standard normal
mixture distribution with the maximum number of components $K=10$. The
number of estimated normal components {\r based on the BIC} is
reported in the column \comment{\texttt{Mclust}}. %\emph{Mclust}.
Then the selected components are combined hierarchically to clusters
by calling function \texttt{clustCombi} from the same package (column
\texttt{clustCombi}). %\emph{clustCombi}).
The number of clusters is chosen by visual detection of the change
point in the plot of the rescaled differences between successive
entropy values, as suggested by
\cite{Mix:BaudryRafteryCeleuxLoGottardo2010}.  Furthermore, to compare
our results to those obtained if a cluster distribution is modeled by
a single normal distribution only, a sparse finite mixture model
\comment{with $K=10$} \citep{Mix:MalsinerWalliFruehwirthGruen2014} is
fitted to the data sets (column \emph{SparseMix}).  The results of
fitting a sparse hierarchical mixture of mixtures model \comment{with
  $K=10$} %, $L=4$, and $L=5$ to the data
are given in column \emph{SparseMixMix}, \comment{where $L=5$ is
  compared to our default choice of $L=4$ to investigate robustness
  with respect to {\bl the choice of} $L$.}
% The results for different values of $L$ are compared to investigate the robustness of the recommended default choice of $L=4$.}
For each estimation, MCMC sampling is run for 4000 iterations after a burn-in of 4000
iterations.

As can be seen in Table~\ref{table:benchmarkNEW}, for all data
sets %we are
\comment{the sparse hierarchical mixture of mixtures model is} able to
capture the data clusters quite well both in terms of the estimated
number of clusters and the clustering quality measured by the
misclassification rate as well as the adjusted Rand index. In general,
our approach is not only outperforming the standard model-based
clustering model using mixtures of Gaussians regarding both measures,
but also the approach proposed by
\cite{Mix:BaudryRafteryCeleuxLoGottardo2010}.  In addition, it can be
noted that for \commentG{all data sets the estimation results remain quite stable,
 if the number of subcomponents $L$ is
increased to $5$,} see the last
column in Table~\ref{table:benchmarkNEW}.  The results for the Yeast
data set are of particular interest as they indicate that
\texttt{clustCombi} completely fails. Although the misclassification
rate of 25\% implies that only a quarter of the observations is
assigned to ``wrong'' clusters, inspection of the clustering obtained
\commentG{reveals that} %indicates that in the solution obtained
almost all observations are lumped together in
% one very
\commentG{a single, very}
large cluster, whereas the few remaining
observations are split into five very small clusters. This bad
clustering quality is better reflected by the adjusted Rand index
which takes a negative value ($\mathit{adj}=-0.02$), i.e.~is ``worse
than would be expected by guessing'' %, as noted by
\commentG{\citep{Mix:FranczakBrowneMcNicholas2012}}. For the flower data set,
{\r more} \commentG{results} are given in Appendix~D where the obtained
clustering and cluster distributions are illustrated.

\begin{table}[t!]
%\begin{sidewaystable}
\centering
\begin{footnotesize}
\begin{tabular}{@{}l@{}ccc|ll|l|ll}
  \hline
  \multicolumn{4}{c}{}& \multicolumn{2}{|c|}{\emph{Mclust}}& \multicolumn{1}{|c|}{\emph{SparseMix}}&\multicolumn{2}{|c}{\emph{SparseMixMix}}\\
  \multicolumn{4}{c|}{}& \multicolumn{2}{|c|}{\comment{$K=10$}}& %{$K^{max}=10$}&
  \multicolumn{1}{|c|}{$K=10$}&\multicolumn{2}{|c}{$K=10$}\\
  \hline
  %Data set & $N$ &  $r$ &  $K^{true}$ &  \texttt{Mclust} &  \texttt{clustCombi }&   &\multicolumn{2}{|c}{}\\ &  &   &     &   & &$L=1$& $L=4$&$L=5$\\
    Data set & $N$ &  $r$ &  $K^{true}$ &  \texttt{Mclust} &  \texttt{clustCombi }&  \comment{$L=1$}& \comment{$L=4$} & \comment{$L=5$}\\
  \hline
  Yeast  &626   &3   &2   &$8$ \emph{(.50, .20)}   &$6$ \emph{(-.02, 0.25)}  &$6$ \emph{(.48, .23)}      &$\mathbf{2}$ \emph{(.68, .08)} &$2$ \emph{(.71, .07) }  \\
%  &      &   &   &$\mathit{adj}=.50$   &$\mathit{adj}=-.0$2&$\mathit{adj}=.48$     &$\mathit{adj}=.68$  & $\mathit{adj}=.71$    \\
%  &      &   &   &$\mathit{er}=.20$   &$\mathit{er}=.25$&$\mathit{er}=.23$    &$\mathit{er}=.08$ &$\mathit{er}=.07$    \\
%  \hline\emph{
  Flea beetles\phantom{1}  &74   & 6  &  3 &$5$ \emph{(.77, .18)}  & $4$ \emph{(.97, .03)} &$3$ \emph{(1.00, .00)}    &$\mathbf{3}$ \emph{(1.00, .00)} & $3$  \emph{(1, .00)}  \\
%  beetles &     &   &   &$\mathit{adj}=.77$   &$\mathit{adj}=.97$  &$\mathit{adj}=1$   & $\mathit{adj}=1$ &$\mathit{adj}=1$    \\
%  &     &   &   &$\mathit{er}=.18$   &$\mathit{er}=.03$   & $\mathit{er}=.00$ &$\mathit{er}=.00$  & $\mathit{er}=.00$  \\
%  \hline
  AIS   &202   &3   & 2  &$3$ \emph{(.73, .13)}  &$2$ \emph{(.66, .09)}&$3$ \emph{(.76, .11)}   & $\mathbf{2}$ \emph{(.81, .05})& $2$  \emph{(.76, .06)}\\
%  &      &   &   &$\mathit{adj}=.73$   &$\mathit{adj}=.66$ &$\mathit{adj}=.76$   &$\mathit{adj}=.81$ &$\mathit{adj}=.76$\\
%  &     &   &   &$\mathit{er}=.13$   &$\mathit{er}=.09$ & $\mathit{er}=.11$   &$\mathit{er}=.05$&  $\mathit{er}=.06$  \\
%  \hline
  Wisconsin&569   & 3  & 2  &$4$ \emph{(.55, .30)}  &$4$ \emph{(.55, .30})&$4$ \emph{ (.62, .21)}    &$\mathbf{2}$ \emph{(.82, .05)} &$2$  \emph{(.82, .05)}   \\
%  &      &   &   &$\mathit{adj}=.55$   &$\mathit{adj}=.55$  &$\mathit{adj}=.62$    & $\mathit{adj}=.82$&$\mathit{adj}=.82$   \\
%  &      &   &   &$\mathit{er}=.30$   &$\mathit{er}=.30$    &$\mathit{er}=.21$    & $\mathit{er}=.05$&$\mathit{er}=.05$     \\
%    \hline
  Flower&400   & 2  & 4  &$6$ \emph{(.52, .35)}  &$4$ \emph{(.99, .01)}&$5$ \emph{ (.67, .20)}    &$\mathbf{4}$ \emph{(.97, .01)}& $4$  \emph{ (.97, .02)}  \\
%    &      &   &   &$\mathit{adj}=0.52$   &$\mathit{adj}=0.99$  &$\mathit{adj}=0.67 $    & $\mathit{adj}=0.97$&$\mathit{adj}=0.97$   \\
%  &      &   &     &$\mathit{er}=0.35$   &$\mathit{er}=0.01$     &$\mathit{er}=0.20 $    &  $\mathit{er}=0.01$  &$\mathit{er}=0.02$     \\
%  \hline
%  Faithful  &272   &2   & $-$  &$3$   &$2$ &$3$    &$\mathbf{2}$ &$2$   \\
%  &      &   &   &$-$  &  $-$&  $-$& $-$  & $-$  \\
%  &      &   &   &$-$  &  $-$&  $-$& $-$  & $-$  \\
  \hline
\end{tabular}
\end{footnotesize}
\caption{\footnotesize Results for the estimated number of data
  clusters $\hat{K}_0$ for various benchmark data sets, using the functions
  \texttt{Mclust} to fit a standard mixture model \comment{with $K=10$} and
  \texttt{clustCombi} to estimate a mixture with combined components (column \emph{Mclust}), using a sparse finite mixture model \comment{with $K=10$}
  (column \emph{SparseMix}), and estimating a sparse hierarchical mixture of   mixtures model   \comment{with $K=10$, % clusters
    $\phi_B=0.5$ and $\phi_W=0.1$, and  $L=4,5$ (column \emph{SparseMixMix}). Priors and hyperparameter
    specifications are selected as described in {\r Section~\ref{sec:model-specification}}.
    In parentheses, the  adjusted Rand index  (\lq\lq 1\rq\rq\ corresponds to perfect classification) and the proportion of misclassified observations (\lq\lq 0\rq\rq\ corresponds to perfect classification) are reported.}}
  % $\mathit{adj}$ reports the adjusted Rand index (\lq\lq 1\rq\rq\
  % corresponds to perfect classification)
  % and $\mathit{er}$ gives the proportion of misclassified
  % observations \comment{(\lq\lq 0\rq\rq\ corresponds to perfect
  % classification)}.}
\label{table:benchmarkNEW}
%\end{sidewaystable}
\end{table}

\begin{figure}[t!]
\centering
    \includegraphics[width=0.35\textwidth, trim = 0 10 0 78, clip]{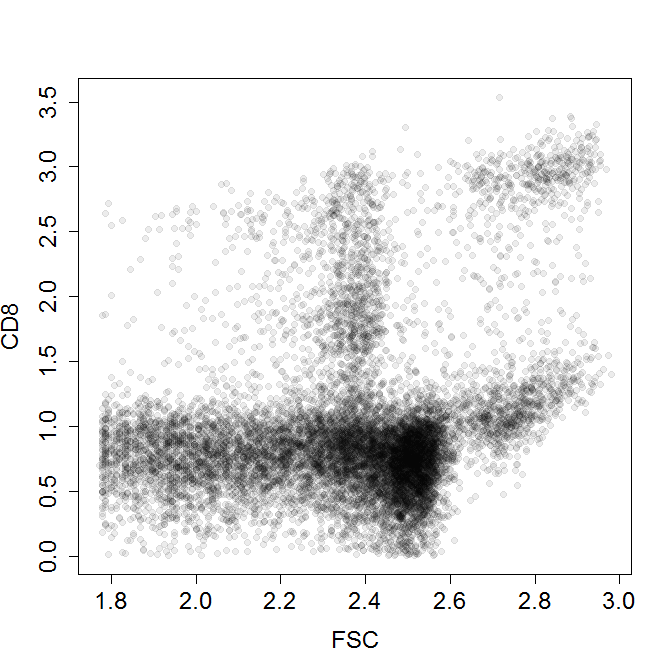}
    \includegraphics[width=0.42\textwidth, trim = 0 10 0 135, clip]{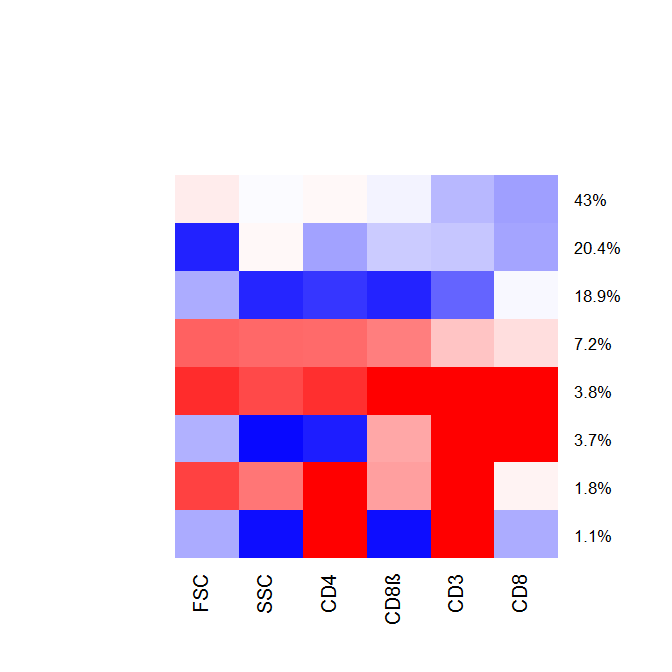}
    \caption{\footnotesize Flow cytometry data set GvHD. Scatter plot
      of two variables (``FSC'', ``CD8'') (left-hand side), and
      heatmap of the clustering results by fitting a sparse
      hierarchical mixture of mixtures model (right-hand side). In the
      heatmap, each row represents the location of a six-dimensional
      cluster, and each column represents a particular marker. The
      red, white and blue colors denote high, medium and low
      expression, respectively.}
  \label{plot:pyne}
\end{figure}

  In order to investigate the performance of our approach on  larger
  data sets with a slightly different cluster structure, we fit the sparse
  hierarchical mixture of mixtures model to two flow cytometry data sets.
  % analyzed, %first by \cite{Mix:Brinkman2007} and then,
  % among others, by \cite{Mix:FruehwirthPyne2010} using mixtures.
  These applications also allow us to indicate how the prior settings need to
  be adapted if a different cluster structure is assumed to be present
  in the data.  As generally known,
  flow cytometry data exhibit non-Gaussian characteristics such as
  skewness, multimodality and \commentG{a} large number of
  outliers, as can be seen in the scatter plot
  of two variables of the GvHD data set in  Figure~\ref{plot:pyne}.
  Thus,  we specified a sparse hierarchical mixture of mixtures model
  with $K=30$ clusters and increased the number of subcomponents
  forming a cluster to $L=15$ in order to handle more complex shapes
  of the cluster distributions given the large amount of data. Since
  the flow cytometry data clusters have a lot of outliers similar to
  the clusters generated by \comment{shifted asymmetric Laplace
    ($\mathit{SAL}$)} distributions (see Appendix F), we
  {\r substitute the hyperprior $\bC_{0k} \sim \cW_r(g_0,\bG_0)$
    by the fixed value $\bC_{0k}= g_0\bG_0^{-1}$ and set $\lambda_{kj}
    \equiv 1$, {\bl $j=1,\ldots, r$}}
  % omit the hyperprior $C_{0k} \sim \cW_r(g_0,\bG_0)$
  to prevent that within a cluster the subcomponent covariance
  matrices \commentG{are overly shrunken and} become too similar.  In
  this way, subcomponent covariance matrices are allowed to vary
  considerably within a cluster and capture both a dense cluster
  region around the cluster center and scattered regions at the
  boundary of the cluster.

\begin{figure}[t!]
	\centering
	\includegraphics[width=0.325\textwidth, trim = 0 30 0 90, clip]{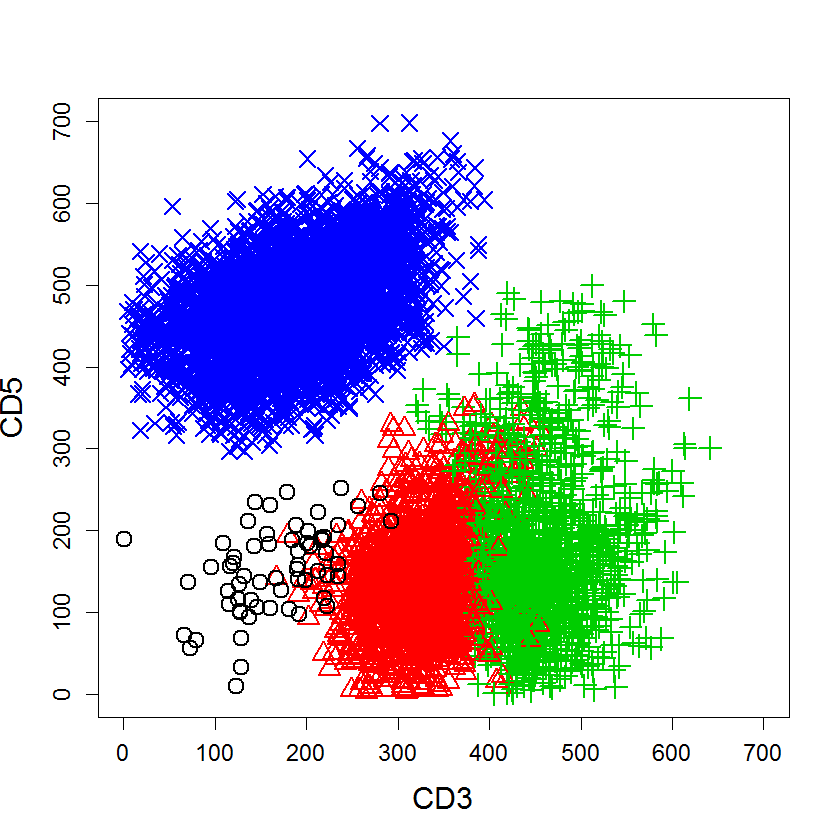}
	\includegraphics[width=0.325\textwidth, trim = 0 30 0 90, clip]{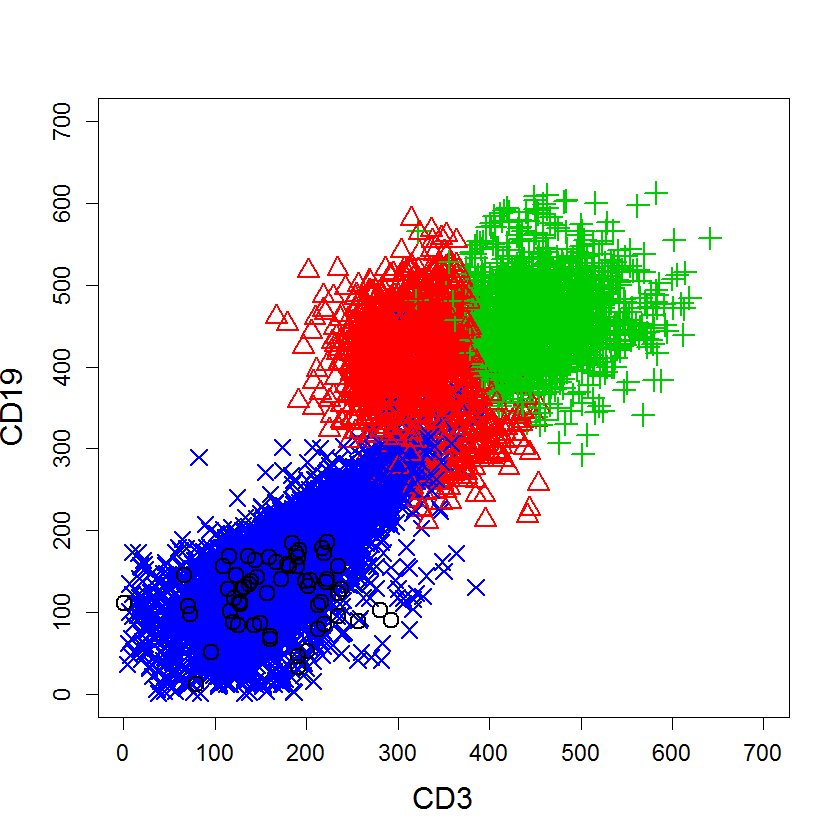}
	\includegraphics[width=0.325\textwidth, trim = 0 30 0 90, clip]{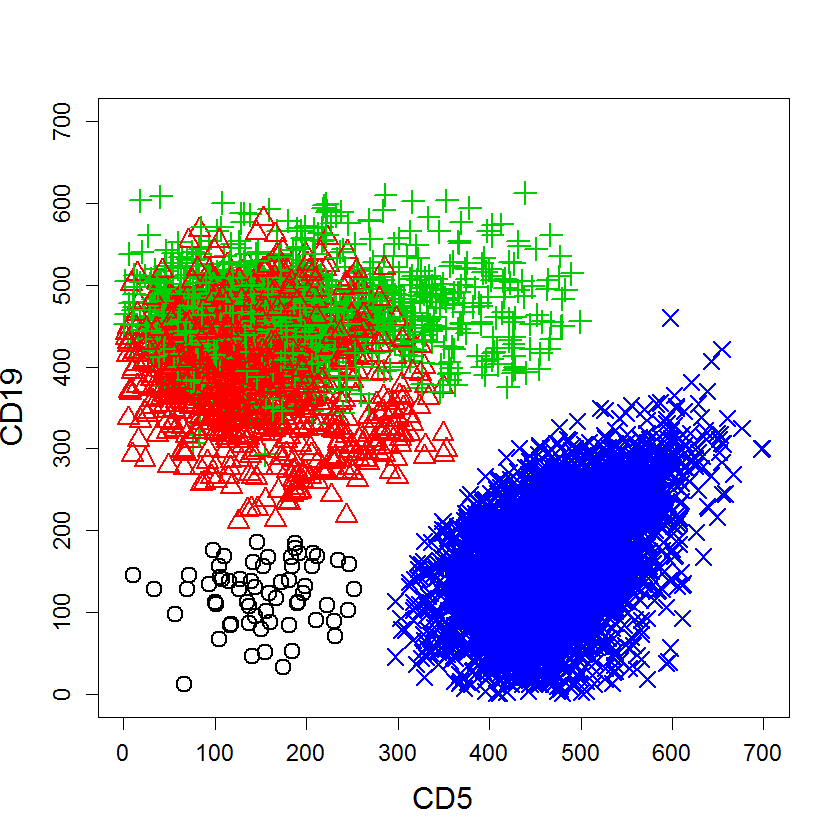}
	\caption{\footnotesize Flow cytometry data set DLBCL. Scatterplot of the clustering results.} \label{plot:DLBCL}
\end{figure}

  We {\bl fit} this sparse hierarchical mixture of mixtures model to the
  DLBCL data after removing 251 dead cells.
  % We removed 251 dead cells and fitted a sparse hierarchical mixture
  % of mixtures model to the remaing 7932 observations by defining the
  % same prior parameter setting as for the GvHD flow cytometry data
  % set.
  For most MCMC runs after a few hundred iterations all but four
  clusters become empty during MCMC sampling. The estimated four
  cluster solution coincides almost exactly with the cluster solution
  obtained with manual gating; the adjusted Rand index is 0.95 and the
  error rate equals 0.03. This error rate outperforms the error rate
  of 0.056 reported by \cite{Mix:LeeMcLachlan2013b}. In
  Figure~\ref{plot:DLBCL} the estimated four cluster solution is
  visualized.

When fitting a sparse hierarchical mixture of mixtures model to the GvHD data,
the classifications resulting from different runs of the MCMC algorithm seemed to be rather stable.
The obtained solutions differ mainly in the size of the two
large clusters with low expressions. These, however, are supposed to
not contain any information regarding the development of the disease.
 \commentG{On the right hand side of Figure~\ref{plot:pyne},} the results of one specific run are
shown in a heatmap. In this run, we found eight clusters which are
similar to those reported by \cite{Mix:FruehwirthPyne2010} when
fitting a skew-$t$ mixture model \commentG{to these data}. In the heatmap each row represents
the location of a six-dimensional cluster, and each column represents
a particular marker (variable). The red, white and blue colors denote
high, medium and low expressions.

As in \cite{Mix:FruehwirthPyne2010}, we identified two larger clusters
(43\% and 20.4\%, first two rows in the heatmap) with rather low
expressions in the last four variables.
%The first two lines represent the location means of the two large clusters with low expressions.
We also identified a smaller cluster (3.8\%, forth row from the
bottom) representing live cells (high values in the first two
variables) with a unique signature in the other four variables (high
values in all four variables).
% In the fourth row from the bottom clearly a small cluster consisting
% of a high expression in all 6 dimensions can be identified.  This
% clis in line with the results reported by \cite{Mix:Brinkman2007}
% and \cite{Mix:FruehwirthPyne2010}.
Also two other small clusters can be identified (second and third row
from the bottom) which have a signature very similar to the clusters
found by \cite{Mix:FruehwirthPyne2010}, and thus our results confirm
their findings.

\section{Discussion}\label{sec:discussion}

We propose suitable priors for fitting an identified mixture of normal
mixtures model within the Bayesian framework of model-based
clustering. This approach allows for (1) automatic determination of
the number of clusters and (2) semi-parametric approximation of
non-Gaussian cluster distributions by mixtures of normals. We only
require the assumption that the cluster distributions are dense and
connected.  Our approach consists in the specification of structured
informative priors on all model parameters.   \comment{This imposes a rigid hierarchical structure on the normal subcomponents {\bl and} allows for simultaneous estimation of the number of clusters and their approximating distributions}.
%Through the prior specification, a rigid hierarchical structure on the normal subcomponents is imposed which allows for simultaneous estimation of the number of clusters and their approximating distributions.
 This is in contrast to the two-step merging approaches, where in the first
step the data distribution is approximated by a suitable normal
mixture model.  However, because this approximation is {\bl made} without
{\bl taking}  the data clusters into account which are reconstructed
only in the second step of the procedure, the general cluster
structure might be missed by these approaches.

As we noted in our simulation studies, the way in which the cluster
mixture distributions are modeled by the subcomponent densities is
crucial for the clustering result.  Enforcing overlapping subcomponent
densities is essential in order to avoid that a single subcomponent
becomes too narrow thus leading to a small {\bl a~posteriori} cluster
probability for observations from this subcomponent. {\r Also,
  enforcing {\bl that observations are assigned to \emph{all} subcomponents}
  during MCMC sampling is important as the estimation of empty
  subcomponents would bias the resulting cluster distribution because
  of the ``prior'' subcomponents.
}% \footnote{\commentG{Bei diesem Satz verstehe ich leider nicht mehr, was gemeint war.}}
For modeling large, overlapping subcomponent densities, crucial model
parameters are the a~priori specified covariance matrix of the
subcomponent means and the scale matrix of the inverse Wishart prior
for the subcomponent covariance matrices.  We select both crucial
hyperparameters based on %considerations resulting from
the variance decomposition of a mixture of mixtures model.

We found a prior setting which is able to capture dense and connected
data clusters in a range of benchmark data sets.  However, if interest
lies in detection of different cluster shapes, a different tuning
of the prior parameters may be required.  Therefore, it would be
interesting to investigate in more detail how we can use certain prior
settings to estimate certain kinds of data clusters. Then it would be
possible to give recommendations which prior settings have to be used
in order to capture certain types of data clusters. For instance,
mixtures of shifted asymmetric Laplace ($\mathit{SAL}$) distributions,
introduced by \cite{Mix:FranczakBrowneMcNicholas2012}, have cluster
distributions which are non-dense and have a strongly asymmetric shape
with comet-like tails. In this case, the prior specifications given in
Section~\ref{sec:model-specification} are not able to capture the
clusters \comment{and need to be tuned to capture also this special
  kind of data clusters, see the example given in Appendix~F.}
%. However, they can be tuned in such a way to capture also this special kind of data clusters, as can be seen in the example given in Appendix~E.

Although our approach to estimate the number of clusters worked well
for many data sets, we encountered mixing problems with the blocked
conditional Gibbs sampler outlined in Appendix~A, in particular in
high dimensional spaces with large data sets.  To alleviate this
problem, a collapsed sampler similar to \citet{fal-bar:gib} could be
derived for finite mixtures. However, we leave this for future
research.
% {\r In addition, some adaptations of the prior settings might be
% required in order to take into account that more complex shapes of
% the cluster distributions can be approximated if more data is
% available.}

\bigskip
\begin{center}
{\bf SUPPLEMENTARY MATERIAL}
\end{center}

%\begin{description}
%\item [Appendix:] The appendix contains
%\begin{description}
\noindent \underline{\textsc{Appendix}} containing (A) the MCMC scheme
to estimate a mixture of mixtures model, (B) a detailed description of
the post-processing strategy based on the point process
representation, (C) {{\bl the} simulation studies described in Section 4},
(D) a description of the data sets studied in
Section~\ref{sec:applications}, (E) issues with the merging approach,
and (F) estimation of data clusters generated by a $SAL$-distribution
\citep{Mix:FranczakBrowneMcNicholas2012}. (Appendix.pdf)

%%\item[R code:] The R code
\noindent \underline{\textsc{R code}} implementing the sparse hierarchical mixture of  mixtures model (Code.zip).
%%\end{description}

\spacingset{1.45} % DON'T change the spacing!

\appendix
\renewcommand\thefigure{\thesection.\arabic{figure}}
\renewcommand\thetable{\thesection.\arabic{table}}

\section{MCMC sampling scheme}\label{app:append-mcmc-sampl}
Estimation of a sparse hierarchical mixture of mixtures model is performed through MCMC
sampling based on data augmentation and Gibbs sampling.  To indicate
the cluster to which each observation belongs, latent allocation
variables $\bS=(S_1,\ldots,S_N)$ taking values in $\{1,\ldots,K\}^N$
are introduced such that
\begin{equation*}
p(\by_i|\btheta_1,\ldots,\btheta_K, S_i=k)=p_{k} (\by_i|\btheta_k), \quad \text{ and } \quad  \Prob{S_i=k|\boldeta}=\eta_k.
\end{equation*}
Additionally, to indicate the subcomponent to which an observation
within a cluster is assigned to, latent allocation variables
$\bI=(I_1,\ldots,I_N)$ taking values in $\{1,\ldots,L\}^N$ are
introduced such that
\begin{equation*}
p_k(\by_i|\btheta_k,S_i=k,I_i=l)=f_{\cN} (\by_i|\bmu_{kl},\bSigma_{kl}) \quad \text{ and } \quad \Prob{I_i=l|S_i=k,\bw_k}=w_{kl}.
\end{equation*}
Based on the priors specified in Section {2.2}, with fixed
hyperparameters $(e_0,d_0,c_0,g_0,\bG_0,\bB_0,\bm_0,\bM_0,\nu)$, the
latent variables $(\bS, \bI)$ and parameters
$(\boldeta,\bw_k,\bmu_{kl},\bSigma_{kl},$ $\bC_{0k},$ $\bb_{0k},$
$\lambda_{kj})$, $k=1,\dots,K$, $l=1,\ldots,L$, $j=1,\ldots,r$, are
sampled from the posterior distribution using the following Gibbs
sampling scheme. Note that the conditional distributions given do not
indicate that conditioning is also on the fixed hyperparameters.
\begin{enumerate}[(1)]
	\item Sampling steps on the  level of the cluster distribution:
	\begin{enumerate}
		\item \emph{Parameter simulation step} conditional on the
		classifications $\bS$.  Sample $\boldeta | \bS $ from
		$Dir(e_1,\ldots,e_K)$, $e_k=e_0+N_k$, \comment{$k=1,\dots,K$,}
		where $ N_k=\#\{S_i|S_i=k\}$ is the number of observations allocated to cluster
		$k$.
		\item \emph{Classification step} for each observation $\by_i$
		conditional on cluster-specific parameters. For each $i=1,\ldots,N$ sample the cluster assignment $S_i$ from
		\begin{align}  \label{liksiapp}
		\Prob{S_i=k|\by_i,\btheta,\boldeta} &\propto \eta_k p_k(\by_i|\btheta_k), \; k=1,\ldots,K,
		\end{align}
		where  $p_k(\by_i| \btheta_k)$ is the semi-parametric mixture approximation of the cluster density:
		\begin{align*}
		p_k(\by_i| \btheta_k) &= \sum_{l=1}^{L} w_{kl} f_{\cN}( \by_i | \bmu_{kl}, \bSigma_{kl}).
		\end{align*}
		Note that clustering of the observations is performed on the
		upper level of the model, using a collapsed Gibbs step, where the
		latent, within-cluster allocation variables $\bI$ are integrated
		out.
	\end{enumerate}
	
	\item Within each cluster $k$, $k=1,\ldots,K$:
	\begin{enumerate}
		\item \emph{Classification step} for all observations $\by_i$, assigned to cluster $k$ (i.e.~$S_i=k$),
		conditional on  the subcomponent weights
		and the subcomponent-specific parameters. For each $i \in \{i = 1,\ldots,N: S_i = k\}$
		sample $I_i$ from
		\[\Prob{I_i=l| \by_i,\btheta_k,S_i=k} \propto w_{kl}
		f_{\cN}(\by_i|\bmu_{kl},\bSigma_{kl}), \; l=1,\ldots,L.\]
		\item \emph{Parameter simulation step} conditional on the
		classifications $\bI$ and $\bS$:
		\begin{enumerate}
			\item Sample $\bw_k|\bI, \bS$ from $Dir(d_{k1},\ldots,d_{kL})$, $d_{kl}=d_0+N_{kl}$,
			\comment{$l=1,\ldots,L$,}
			where $ N_{kl}=\#\{I_i=l|S_i=k\}$ is the
			number of observations allocated to subcomponent $l$ in cluster
			$k$.
			\item For $l=1,\ldots,L$: Sample $\bSigma_{kl}^{-1}|\bS,\bI,\bmu_{kl},\bC_{0k}, \by \sim
			\cW_r(c_{kl},\bC_{kl})$, where
			\begin{align*}
			c_{kl} &= c_{0}+N_{kl}/2,\\
			\bC_{kl} &=\bC_{0k} +\frac{1}{2}\sum \limits_{i:I_i=l,S_i=k} (\by_i-\bmu_{kl})(\by_i-\bmu_{kl})'.
			\end{align*}
			\item For $l=1,\ldots,L$: Sample $\bmu_{kl}|\bS, \bI, \bb_{0k}, \bSigma_{kl},\bLambda_k, \by \sim \cN_r(\bb_{kl},\bB_{kl})$, where
			\begin{align*}
			\bB_{kl} &=( \tilde \bB_{0k}^{-1}+ N_{kl} \bSigma_{kl}^{-1})^{-1},\\
			\bb_{kl} &= \bB_{kl}(\tilde \bB_{0k}^{-1}\bb_{0k}+ \bSigma_{kl}^{-1} N_{kl} \bar{\by}_{kl}),
			\end{align*}
			with $\tilde \bB_{0k}=\sqrt{\bLambda_k} \bB_{0} \sqrt{\bLambda_k}$, $\bLambda_k=\text{diag}(\lambda_{k1},\ldots,\lambda_{kr}),$ and
			$\bar{\by}_{kl} = 1/N_{kl} \sum \limits_{i:I_i=l,S_i=k} \by_i$
			\comment{being equal to the subcomponent mean for $N_{kl}>0$ and $N_{kl} \bar{\by}_{kl}=0$, otherwise.}
		\end{enumerate}
	\end{enumerate}
	
	\item For each cluster $k$, $k=1,\ldots,K$: Sample the random hyperparameters  $\lambda_{kj}$, $\bC_{0k},\bb_{0k}$ from their full conditionals:
	\begin{enumerate}
		\item For $j=1,\ldots,r$: Sample $\lambda_{kj}| \bb_{0k}, \bmu_{k1}, \ldots, \bmu_{kL} \sim
		\mathcal{GIG}(p_{kL},a_{kj},b_{kj})$, where $\mathcal{GIG}$ is
		the generalized inverted Gaussian distribution and
		\begin{align*}
		p_{kL} &= -L/2 + \nu,\\
		a_{kj} &= 2 \nu,\\
		b_{kj} &= \sum_{l=1}^L (\mu_{kl,j}-b_{0k,j})^2/B_{0,jj}.
		\end{align*}
		\item Sample $\bC_{0k}| \bSigma_{k1}, \ldots, \bSigma_{kL} \sim \cW_r(g_{0}+ Lc_{0},\bG_{0}+\sum_{l=1}^L \bSigma_{kl}^{-1})$.
		\item Sample $\bb_{0k} | \tilde \bB_{0k}, \bmu_{k1}, \ldots, \bmu_{kL}  \sim \cN_r(\tilde{\bm}_k,\tilde{\bM}_k)$, where
		\begin{align*}
		\tilde{\bM}_k &= (\bM_0^{-1}+ L \tilde \bB_{0k}^{-1})^{-1},\\
		\tilde{\bm}_k &= \tilde{\bM}_k
		\left(\bM_0^{-1}\bm_0+ \tilde \bB_{0k}^{-1}\sum_{l=1}^L
		\bmu_{kl}\right).
		\end{align*}
	\end{enumerate}
\end{enumerate}

\newpage

\section{Identification through clustering in the point process
	representation} \label{app:clustering}

{\bl Various} post-processing approaches have been proposed for the MCMC
output of finite or infinite mixture models (see, for example,
\citealt{mol-etal:bay} or
\citealt{Mix:Jasra+Holmes+Stephens:2005}). We pursue an approach
which aims at determining a unique labeling of the MCMC draws after
selecting a suitable number of clusters in order to base any
posterior inference on the relabeled draws, such as {\r for example}
the determination of cluster assignments.

To obtain a unique labeling of the clusters,
\cite{Mix:Fruehwirth2006} suggested to post-process the MCMC output
by clustering \commentG{a vector-valued functional $f(\btheta_k)$  of the cluster-specific
	parameters $\btheta_k$}
% draws $\{\btheta_1,\ldots,\btheta_K \}$
in the point process
representation.  The point process representation has the advantage
that it allows to study the posterior distribution of
cluster-specific parameters regardless of potential label switching,
which makes it very useful for identification.

If the number $K$ of components matches the true number of clusters,
it can be expected that the vector-valued functionals of the
posterior draws cluster around the ``true'' points
$\{f(\btheta_1),\ldots,f(\btheta_K)\}$
\citep[p.~96]{Mix:Fruehwirth2006}. However, in the case of an
overfitting mixture where draws are sampled from empty components,
the clustering procedure has to be adapted as suggested in
\cite{Mix:Fruehwirth2011} and described in more details in
\cite{Mix:MalsinerWalliFruehwirthGruen2014}. Subsequently, we
describe how this approach can be applied to identify
cluster-specific characteristics for the sparse hierarchical mixture of
mixtures model.

First, we estimate the number of non-empty clusters $\hat{K}_0$ on the upper level of
the sparse hierarchical mixture of mixtures model. For
this purpose, during MCMC sampling for each iteration $m$ the number
of non-empty clusters $K_0^{(m)}$ is determined, i.e.~the number of
clusters to which observations have been assigned for this particular
sweep of the sampler:
\begin{equation}
K_0^{(m)}=K-\sum \limits_{k=1}^K I\{N_k^{(m)}=0\}, \label{eq:K0}
\end{equation}
where $N_k^{(m)} =\sum_{i=1} ^N I\{S_i^{(m)}=k\} $ is the number of
observations allocated to cluster $k$ in the upper level of the
mixture for iteration $m$ and $I$ denotes the indicator function.
Then, following \cite{Mix:Nobile2004} we obtain the posterior
distribution of the number $K_0$ of non-empty clusters
$\Prob{K_0=h|\by}, h=1,\ldots, K,$ on the upper level from the MCMC
output.  An estimator of the true number of clusters $\hat{K}_0$ is
then given by the value visited most often by the MCMC procedure,
i.e.~the mode of the (estimated) posterior distribution
$\Prob{K_0=h|\by}$.

After having estimated the number of non-empty clusters $\hat{K}_0$,
we condition the subsequent analysis on a model with $\hat{K}_0$
clusters by removing all draws generated in iterations where the
number of non-empty clusters does not correspond to $\hat{K}_0$. Among
the remaining \comment{$M_0$} draws, only the non-empty clusters are relevant. Hence,
we remove all cluster-specific draws $\btheta_k$ for empty clusters
(which have been sampled from the prior).  The cluster-specific draws
left are samples from $\hat{K}_0$ non-empty clusters and form the
basis for clustering the vector-valued functionals of the draws in the
point process representation into $\hat{K}_0$ groups.

It should be noted, that using only vector-valued functionals of the
unique parameters $\btheta_k$ for this clustering procedure has two
advantages. First, $\btheta_k$ is a fairly high-dimensional parameter
of dimension $d=L-1+Lr(r+3)/2$, in particular if $r$ is large, and the
vector-valued functional allows to consider a lower dimensional
problem \citep[see also][]{Mix:Fruehwirth2006,Mix:Fruehwirth2011}.  In
addition, we need to solve the label switching issue only on the upper
level of the sparse hierarchical mixture of mixtures model. Thus, we
choose vector-valued functionals of the cluster-specific parameters
$\btheta_k$ that are invariant to label switching on the lower level
of the mixture for clustering in the point process representation of
the upper level. We found it {\bl particularly} useful to consider the cluster
means on the upper level mixture, defined by
$\bmu_k^{(m)}=\sum_{l=1}^L w_{kl}^{(m)} \bmu_{kl}^{(m)}$.

Clustering the cluster means in the point process representation
results in a classification sequence $\rho^{(m)}$ for each MCMC
iteration $m$ indicating to which class a single cluster-specific draw
belongs. {\r For this,} \commentG{any clustering algorithm} could be
used, e.g., $K$-means \citep{mix:Hartigan+Wong:1979} or $K$-centroids
cluster analysis \citep{Mix:Leisch2006} where the distance between a
point and a cluster is determined by the Mahalanobis distance, see
\citet[Section~4.2]{Mix:MalsinerWalliFruehwirthGruen2014} for more
details. {\r Only} the classification sequences $\rho^{(m)}$ which
correspond to permutations of $(1,\ldots,\hat{K}_0)$ are used to
relabel the draws. To illustrate this step, consider for instance,
that for $\hat{K}_0=4$, for iteration $m$ a classification sequence
$\rho^{(m)}=(1,3,4,2)$ is obtained through the clustering
procedure. That means that the draw of the first cluster was assigned
to class one, the draw of the second cluster was assigned to class
three and so on.  In this case, the draws of this iteration are
assigned to different classes, which allows to relabel these draws.
As already observed by \cite{Mix:Fruehwirth2006}, all classification
sequences $\rho^{(m)}$, $m=1,\ldots,M$ obtained in this step are
expected to be permutations, if the point process representation of
the MCMC draws contains well-separated simulation clusters.

Nevertheless, it might happen that some of the classification
sequences $\rho^{(m)}$ are not permutations.  E.g., if the
classification sequence $\rho^{(m)}=(3,1,2,1)$ is obtained, then draws
sampled from two different clusters are assigned to the same class and
no unique labels can be assigned.  If only a small fraction
$M_{0, \rho}$ of non-permutations is present, then the posterior draws
corresponding to the non-permutation sequences are removed from the
$M_0$ \comment{draws with  $\hat{K}_0$ non-empty clusters.} %iterations.
For the remaining $M_0 (1 - M_{0, \rho})$ draws, a
unique labeling is achieved by relabeling the clusters according to
the classification sequences \commentG{$\rho^{(m)}$}.  If the fraction $M_{0, \rho}$ is high,
this indicates that in the point process representation clusters are
overlapping. This typically happens if the selected mixture model with
$\hat{K}_0$ clusters is overfitting, see \cite{Mix:Fruehwirth2011}.

This post-processing strategy of the MCMC draws
obtained using the sampling strategy described in
Appendix~\ref{app:append-mcmc-sampl} can be summarized as follows:
\begin{enumerate}
	\item For each iteration $m=1,\ldots,M$ of the MCMC run, determine the
	number of non-empty clusters $K_0^{(m)}$ according to (\ref{eq:K0}).
	
	\item Estimate the number of non-empty clusters by
	$\hat{K}_0=mode(K_0^{(m)})$ as the value of the number of non-empty
	clusters occurring most often during MCMC sampling.

	\item Consider only the \commentG{subsequence %$M_0$
		of all MCMC iterations} {\r of length $M_0$} where
	the number of non-empty clusters $K_0^{(m)}$ is exactly equal to
	$\hat{K}_0$. For each of \commentG{the resulting $m=1, \ldots, M_0$} draws, \commentG{relabel the posteriors draws  $\btheta_1^{(m)}, \ldots, \btheta_K^{(m)}$,
		the weight distribution $\eta_1^{(m)},\ldots, \eta_K^{(m)}$, as well as  the upper level classifications $S_1^{(m)},\ldots,S_N^{(m)}$ such that empty clusters, i.e. clusters with   $N_k^{(m)}=0$, appear last.
		%For each $m=1, \ldots, M_0$,
		Remove the empty clusters and  keep only the draws $\btheta_1^{(m)}, \ldots, \btheta_{\hat{K}_0}^{(m)}$ of the $\hat{K}_0$ non-empty clusters.}
	
	\item \commentG{Arrange the $\hat{K}_0$ cluster means  $\bmu_1^{(m)}, \ldots, \bmu_{\hat{K}_0}^{(m)}$ for all $M_0$ draws %.  Arrange these draws
		in a \lq\lq data matrix\rq\rq\ with
		$\hat{K}_0 \cdot M_0$ rows and $r$ columns such that the first $\hat{K}_0$ rows correspond to the first draw $\bmu_1^{(1)}, \ldots, \bmu_{\hat{K}_0}^{(1)}$, the next $\hat{K}_0$ rows correspond to the second draw $\bmu_1^{(2)}, \ldots, \bmu_{\hat{K}_0}^{(2)}$, and so on. The columns correspond to the different dimensions of  $\bmu$.}
	Cluster all
	$\hat{K}_0 \cdot M_0$ draws into $\hat{K}_0$ clusters using either
	$K$-means \citep{mix:Hartigan+Wong:1979} or $K$-centroids cluster
	analysis \citep{Mix:Leisch2006}.  Either of these cluster algorithms results in a
	classification \commentG{index for each of the $\hat{K}_0 \cdot M_0$  rows of the
		\lq\lq data matrix\rq\rq\ constructed from the MCMC draws.
		This classification vector is rearranged in terms of a   sequence of  classifications $\rho^{(m)}$,  $m=1, \ldots, M_0$,
		where each $\rho^{(m)}=(\rho_1^{(m)}, \ldots, \rho_{\hat{K}_0}^{(m)})$ is a vector of length  $\hat{K}_0$, containing
		the classifications for each draw   $\bmu_1^{(m)}, \ldots, \bmu_{\hat{K}_0}^{(m)}$ at iteration $m$.
		Hence, $\rho_k^{(m)}$  indicates for each single draw $\bmu_k^{(m)}$  to which cluster it   belongs.}
	
	\item \comment{For each iteration $m$, $m=1,\ldots,M_0$, %construct a
		%classification sequence $\rho^{(m)}$ of size $\hat{K}_0$ containing
		% the classifications of each draw at iteration $m$ and
		check whether $\rho^{(m)}$ is a permutation of
		$(1,\ldots,\hat{K}_0)$}.  If not, remove the corresponding draws from
	the MCMC subsample of size $M_0$.  The proportion of classification
	sequences of $M_0$ not being a permutation is denoted by
	$M_{0, \rho}$.
	
	\item For the remaining $M_0(1-M_{0,\rho})$ draws, a unique labeling
	is achieved by resorting the entire vectors of draws
	$\{\btheta_1^{(m)},\ldots,\btheta_{\hat{K}_0}^{(m)} \}$ % from all non-empty clusters
	(not only
	$\bmu_1^{(m)}, \ldots, \bmu_{\hat{K}_0}^{(m)}$),  \commentG{the weight distribution $\eta^{(m)}_1,\ldots, \eta^{(m)}_{\hat{K}_0}$,
		as well as  {\bl relabeling} the upper level classifications $S_1^{(m)},\ldots,S_N ^{(m)}$}
	according to the
	classification sequence $\rho^{(m)}$.
\end{enumerate}
Based on the relabeled draws cluster-specific inference is possible.
For instance, a straightforward way to cluster the data is to assign
each observation $\by_i$ to the cluster $\hat{S}_i$ which is visited
most often. Alternatively, each observation $\by_i$ may also be
clustered based on estimating
$t_{ik}=\Prob{S_i=k|\by_i}$. \commentG{An estimate $\hat{t}_{ik}$ of $t_{ik}$ can be obtained  for each
	$ k=1,\ldots,K$, %\hat{K}_0$,
	by averaging over $\Prob{S_i=k|\by_i,\btheta_k^{(m)},\eta_k^{(m)}},$
	given by Equation~(\ref{liksiapp}) using the relabeled draws.}  Each
observation $\by_i$
is then assigned to that cluster which exhibits the maximum posterior
probability, i.e.~$\hat{S}_i$
is defined in such a way that $\hat{t}_{i,\hat{S}_i}=\max_k
\hat{t}_{ik}$.  The closer
$\hat{t}_{i,\hat{S}_i}$
is to {\bl one}, the higher is the segmentation power for observation
$i$.
Furthermore, the clustering quality of the estimated model can also be
assessed based on estimating the posterior expected entropy. The
entropy of a finite mixture model is defined in \citet{Mix:Celeux1996}
and also described in \citet[][p.~28]{Mix:Fruehwirth2006}. Entropy
values close to zero indicate that observations can unambiguously be
assigned to one cluster, whereas large values indicate that
observations have high {\bl a~posteriori} probabilities for not only one,
but several clusters.

\begin{figure}[t!]
	\centering
	\includegraphics[width=0.32\textwidth]{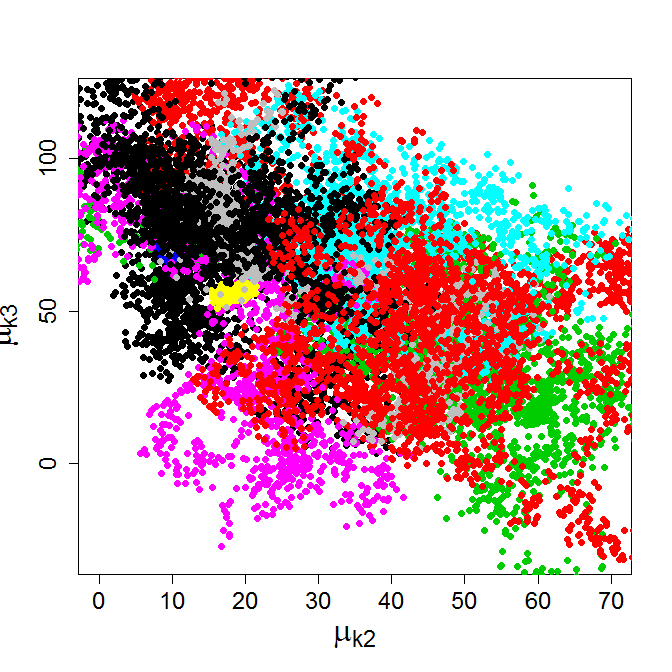}
	\includegraphics[width=0.32\textwidth]{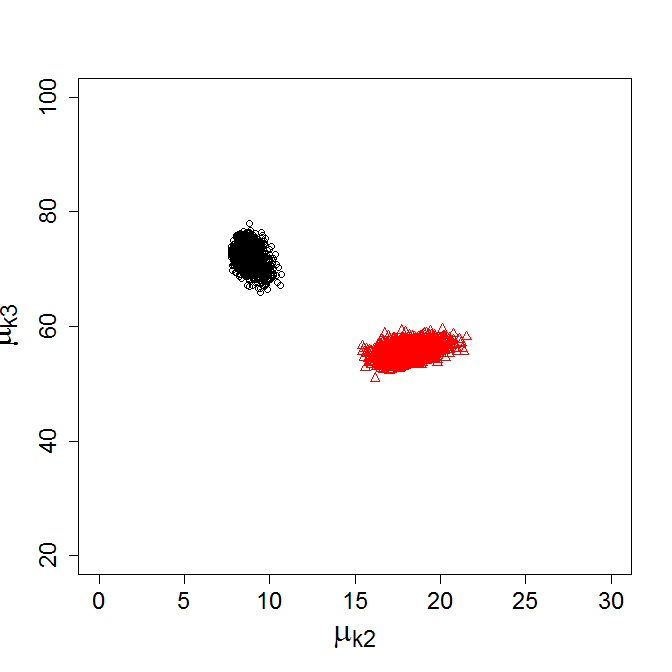}
	\includegraphics[width=0.32\textwidth]{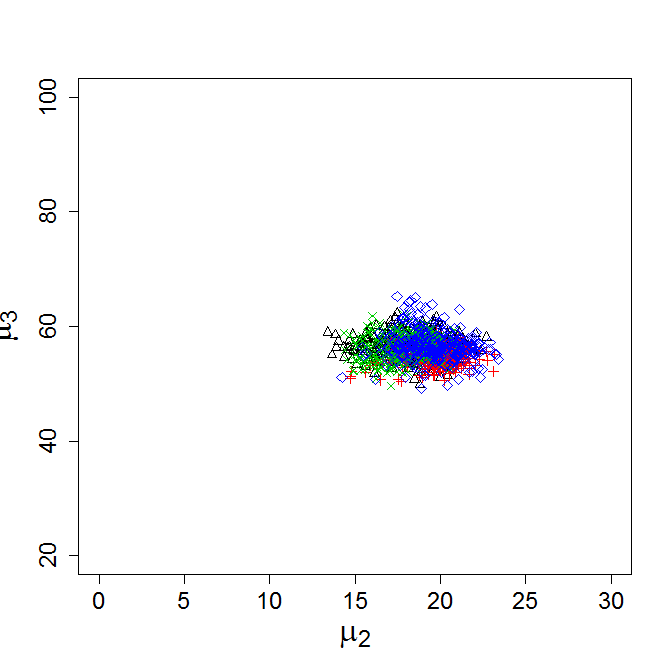}
	\caption{\footnotesize AIS data set, $K=10$, $L=4$, $\phi_B=0.5$,
		$\phi_W=0.1$: Point process representation of the cluster
		means $\bmu_k$ of \emph{all} 10 clusters (left-hand side) and
		only from those where $\hat{K}_0=2$ (middle).  Right-hand side:
		Point process representation of the means of all subcomponents
		forming the cluster in the bottom right in the plot in the
		middle. }\label{plot:ppr1}
\end{figure}

To illustrate identification through clustering the draws in the point
process representation in the present context of a mixture of mixtures
model, a sparse hierarchical mixture of mixtures model with $K=10$
clusters and $L=4$ subcomponents is fitted to the AIS data set (see
Figure~\ref{plot:AisR3} and Section~4). The point process representation of the
weighted cluster mean draws
$\bmu_k^{(m)}=\sum_{l=1}^L w_{kl}^{(m)} \bmu_{kl}^{(m)}$ of \emph{all}
clusters, including empty clusters, is shown in Figure~\ref{plot:ppr1}
on the left-hand side. Since a lot of draws are sampled from empty
clusters, i.e.~from the prior distribution, the plot shows a cloud of
overlapping posterior distributions where no cluster structure can be
distinguished. However, since during MCMC sampling in almost all
iterations only two clusters were non-empty, the estimated number of
clusters is $\hat{K}_0=2$. Thus all draws generated in iterations
where the number of non-empty clusters is different from two and all
draws from empty clusters are removed. The point process
representation of the remaining cluster-specific draws is shown in the
scatter plot in the middle of Figure~\ref{plot:ppr1}. Now the draws
cluster around two well-separated points, and the two clusters can be
easily identified.

To illustrate the subcomponent distributions which are used to
approximate the cluster distribution{\r s}, the point process representation
of the subcomponent means is shown in Figure~\ref{plot:ppr1} on the
right-hand side for the cluster discernible at the bottom right in
Figure~\ref{plot:ppr1} in the middle.  The plot clearly indicates that
all subcomponent means are shrunken toward the cluster mean as the
variation of the subcomponent means is about the same as the variation
of the cluster means.

%\section{Simulation setups}\label{app:simu}

\section{Simulation studies} \label{app:simu} %\label{sec:simulation-studies}

For {\bl both } simulation studies, 10 data sets are generated and a sparse
hierarchical mixture of mixtures model is estimated. Prior
distributions and hyperparameters are specified as described in
Section~2.1 and 2.3.  MCMC sampling is run for
$M=4000$ iterations after a burn-in of $4000$ draws.  For the
sampling, the starting classification of the observations is obtained
by first clustering the observations into $K$ groups using $K$-means
clustering and by then allocating the observations within each group
to the $L$ subcomponents by using $K$-means clustering again.  The
estimated number of clusters is reported in
Tables~\ref{Table:4figures} and~\ref{table:3points}, where in
parentheses the number of data sets for which this number is estimated
is given.

\subsection{Simulation setup I}\label{sec:SimI}

The simulation setup I consists of drawing samples with 800
observations grouped in four clusters. Each cluster is generated by a
normal mixture with a different number of subcomponents.
The four clusters are generated by sampling  from an
eight-component normal mixture with component means
\begin{equation*}\label{}
(\bmu_1 \quad \bmu_2 \quad\ldots \quad \bmu_8)=  \begin{pmatrix} 6 &4&8&22.5&20&22&22&6.5\\ 1.5&6&6&1.5&8&31&31&29\end{pmatrix},
\end{equation*}
variance-covariance matrices
\begin{align*}
\bSigma_1 &=  \begin{pmatrix}4.84 &0\\
0 &2.89\end{pmatrix},&
\bSigma_2&=  \begin{pmatrix} 3.61 & 5.05\\
5.05 &14.44 \end{pmatrix},&
\bSigma_3&=  \begin{pmatrix}3.61 &-5.05\\
-5.05 &14.44 \end{pmatrix},\\
\bSigma_4&=  \begin{pmatrix}  12.25 &0\\
0& 3.24 \end{pmatrix},&
\bSigma_5&=  \begin{pmatrix} 3.24 & 0\\
0 &12.25 \end{pmatrix},&
\bSigma_6&=  \begin{pmatrix} 14.44 &0\\
0& 2.25 \end{pmatrix},\\
\bSigma_7&=  \begin{pmatrix} 2.25 & 0\\
0 &17.64 \end{pmatrix},&
\bSigma_8&=  \begin{pmatrix}2.25  &4.2\\
4.20 &16.0 \end{pmatrix},
\end{align*}
and weight vector $\boldeta=1/4(1/3,1/3,1/3,1/2,1/2,1/2,1/2,1)$.

\begin{figure}[t!]
	\centering
	
	\includegraphics[width=0.32\textwidth]{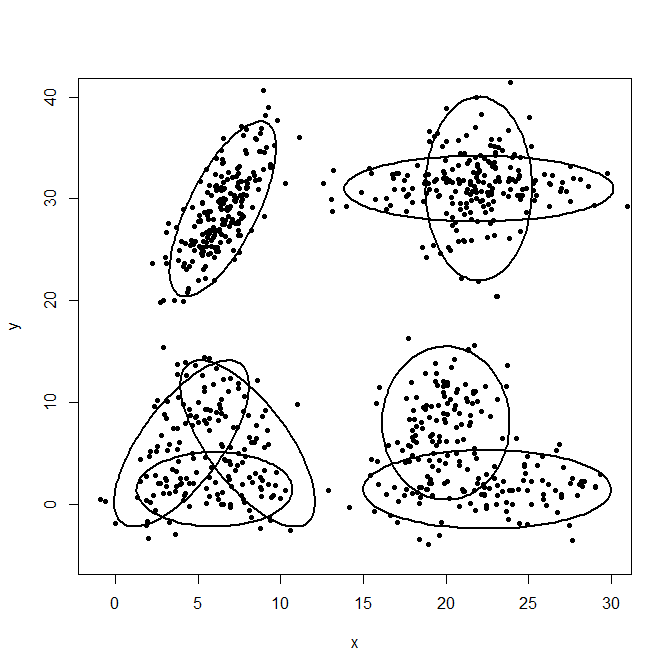}
	\includegraphics[width=0.32\textwidth]{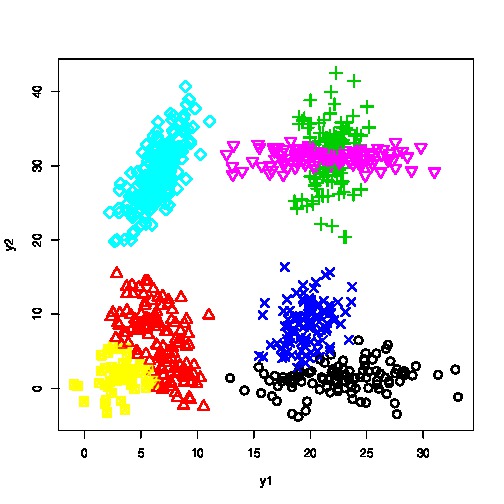}
	\includegraphics[width=0.32\textwidth]{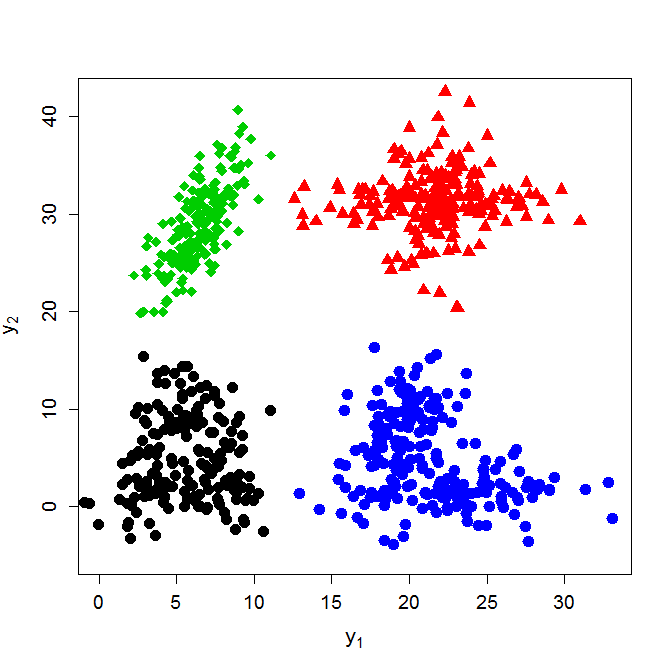}
	\caption{\footnotesize Simulation setup I.  Scatter plot of \comment{one} data
		set with the generating component densities shown with
		90\% probability contour lines (left-hand side), and clustering
		results by estimating a sparse hierarchical mixture of  mixtures model
		with $K=10$, $L=1$ (middle) and $K=10$, $L=4$ (right-hand side).}\label{plot:4figurData}
\end{figure}

In
Figure~\ref{plot:4figurData} the scatter plot of \commentG{one}  data set and the
90\% probability contour lines of the generating subcomponent
distributions are shown. The first three normal distributions generate the triangle-shaped cluster, the next two  the L-shaped cluster, and the last three distributions the cross-shaped and the elliptical cluster.
The number of generating distributions for
each cluster (clockwise from top left) is 1, 2, 2, and 3.  This
simulation setup is inspired by
\cite{Mix:BaudryRafteryCeleuxLoGottardo2010} who use clusters similar
to the elliptical and cross-shaped clusters on the top of the scatter
plot in Figure~\ref{plot:4figurData}.  However, our simulation setup
is expanded by the two clusters at the bottom which have a triangle
and an $L$ shape.  Our aim is to recover the four clusters.

If we estimate a sparse finite mixture model
\citep[see][]{Mix:MalsinerWalliFruehwirthGruen2014}, which can be seen
as a special case of the sparse hierarchical mixture of  mixtures model with number of
subcomponents $L=1$, the estimated number of components is seven, as
can be seen in the classification results shown in
Figure~\ref{plot:4figurData} in the middle plot.  This is to be expected, as
by specifying a standard normal mixture the number of generating
normal distributions is estimated rather than the number of data
clusters.  In contrast, if a sparse hierarchical mixture of mixtures model with
$K=10$ clusters and $L=4$ subcomponents is fitted to the data, all but
four clusters become empty during MCMC sampling and the four data
clusters are captured rather well, as can be seen in the
classification plot in Figure~\ref{plot:4figurData} on the right-hand
side.

In order to study the effect of changing the specified maximum number of clusters $K$
and subcomponents $L$ on the estimation result,  a simulation study
consisting of 10 data sets with the simulation setup as explained
above and varying numbers of clusters $K=4,10,15$ and subcomponents
$L=1,3,4,5$ is performed.  For each combination of $K$ and $L$ the
estimated number of clusters is reported in
Table~\ref{Table:4figures}.

First we study the effect of the number of specified subcomponents $L$
on the estimated number of data clusters.  As can be seen in Table
\ref{Table:4figures}, we are able to identify the true number of
clusters if the number of subcomponents $L$ forming a cluster is at
least three. I.e.~by specifying an overfitting mixture with $K=10$
clusters, for (almost) all data sets superfluous clusters become empty
and using the most frequent number of non-empty clusters as an
estimate for the true number of data clusters gives good results. If a
sparse finite normal mixture is fitted to the data, for almost all
data sets 7 normal components are estimated.  Regarding the maximum
number of clusters $K$ in the overfitting mixture, the estimation
results do scarcely change if this number is increased to $K=15$, as
can be seen in the last row of Table \ref{Table:4figures}. This means
that also in a highly overfitting mixture, all superfluous clusters
become empty during MCMC sampling.

\begin{figure}[t!]
	\centering
	\includegraphics[width=0.32\textwidth, trim = 0 0 0 50, clip]{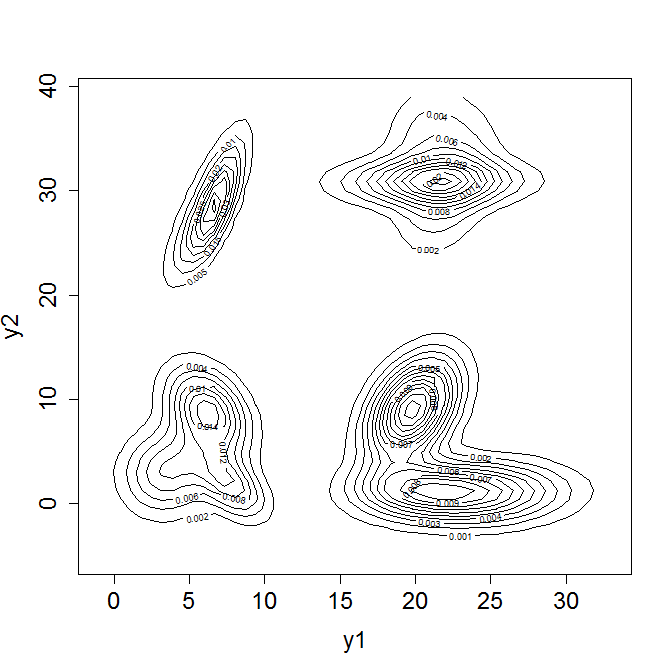}
	\includegraphics[width=0.32\textwidth, trim = 0 0 0 50, clip]{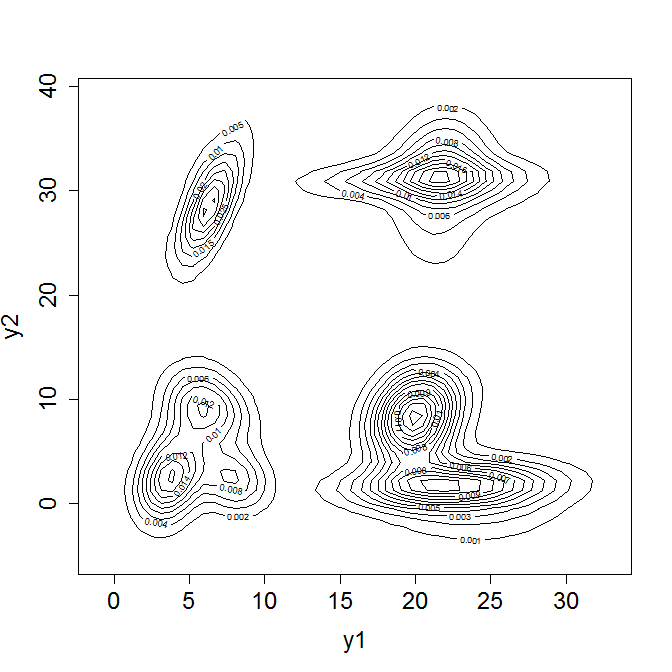}
	\includegraphics[width=0.32\textwidth, trim = 0 0 0 50, clip]{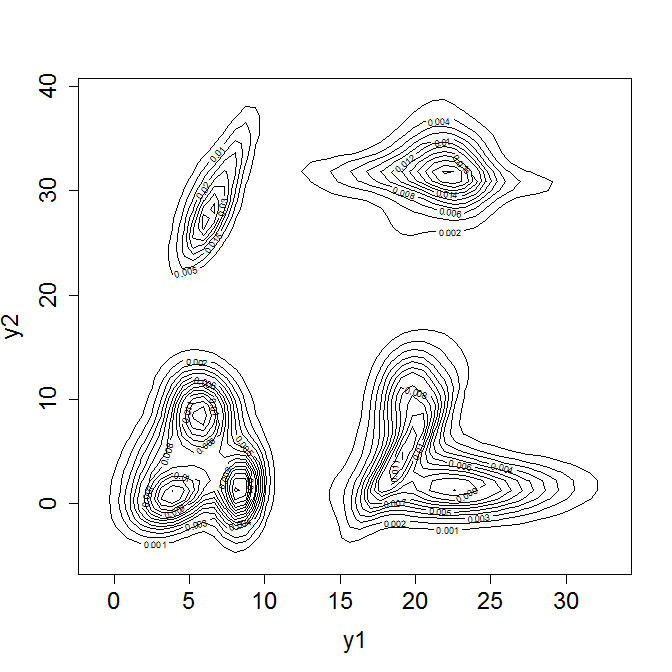}
	\caption{\footnotesize Simulation setup I, $K=10$, {\bl various}
		values of $L$. For the data set in Figure \ref{plot:4figurData},
		the estimated cluster distributions (MAP estimates of means,
		weights, and covariance matrices of the subcomponents) are
		plotted for an increasing number of specified subcomponents
		$L=3,4,5$ (from left to right).} \label{plot:4figureDist}
\end{figure}

In Figure~\ref{plot:4figureDist}, the effect of the number of
subcomponents $L$ on the resulting cluster distributions is
studied. For the data set shown in Figure~\ref{plot:4figurData}, for
an increasing number of subcomponents the estimated cluster
distributions are plotted using the MAP estimates of the weights,
means and covariance matrices of the subcomponents. The estimated
cluster distributions look quite similar, regardless of the size of
$L$.  This \commentG{robustness}  may be due to the smoothing effect of the specified
hyperpriors.

\begin{table}[t!]
	\centering
	\begin{footnotesize}
		\begin{tabular}{|c|cc|cc|cc|cc|}
			\hline
			\backslashbox{$K$}{$L$}& 1&   & 3 & &  4 & & 5 &\\ \hline
			4 & 4(10)&  & {4}(10) & & {4}(10) & &  4(10)& \\
			%     &  & & &  &    &  &  & \\
			\hline
			%  5  &5(9) &    &{4}(10) &  &{4}(10) &  &4(7) &  \\
			%      &4(1)  &    & &  & &  & 5(3) & \\   \hline
			10  &7(9)  &   &{4}(10)  & &{4}(10)  & &4(10) &  \\
			& 6(1)  &  &        &  &                   &  & & \\
			%      &       &   &        &  &                   &  &  & \\
			\hline
			15  &7(9)  &  &{4}(10) & &{4}(9)  & &4(10)  &   \\
			&8(1)  &  & &  & 5(1)&  &  & \\
			\hline
		\end{tabular}
	\end{footnotesize}
	%\caption{\footnotesize  Simulation setup I: Results for the estimated number of non-empty clusters $\hat{K}_0$.   The number of data sets estimating the reported number of non-empty clusters is given in parentheses. %%The number of data sets is 10,
	%$\phi_B=0.5,\phi_W=0.1$, with hyperpriors  $\bC_{0k} \sim \cW_r(g_0,\bG_{0})$ and $\lambda_{kl} \sim \cG(10,10)$}\label{Table:4figures}
	\caption{\footnotesize  \comment{Simulation setup I (based on 10 data sets); true number of clusters equal to 4.
			Results for the estimated number of non-empty clusters $\hat{K}_0$ for various values of $K$ and $L$.
			The number of data sets estimating the reported  $\hat{K}_0$
			is given in parentheses.}
		%$\phi_B=0.5,\phi_W=0.1$, with hyperpriors  $\bC_{0k} \sim \cW_r(g_0,\bG_{0})$ and $\lambda_{kl} \sim \cG(10,10)$
	}\label{Table:4figures}
\end{table}


\subsection{Simulation setup II}\label{sec:simII}
In Section~2.3 it is suggested to specify the
between-cluster variability by $\phi_B=0.5$ and the
between-subcomponent variability by $\phi_W=0.1$. As can be seen in
the previous simulation study in Section~\ref{sec:SimI} this a~priori
specification gives promising results if the data clusters are
well-separated.  However, in contrast to the simulation setup I, in
certain applications data clusters might be close or even
overlapping. In this case, the clustering result might be sensitive in
regard to the specification of $\phi_B$ and $\phi_W$. Therefore, in
the following {\bl simulation study it} is investigated how the
specification of $\phi_B$ and $\phi_W$ affects the identification of
data clusters if they are not well-separated. We want to study how
robust the clustering results are against misspecification of the two
proportions.

\begin{figure}[t!]
	\centering
	\includegraphics[width=0.325\textwidth, trim = 0 20 0 50, clip]{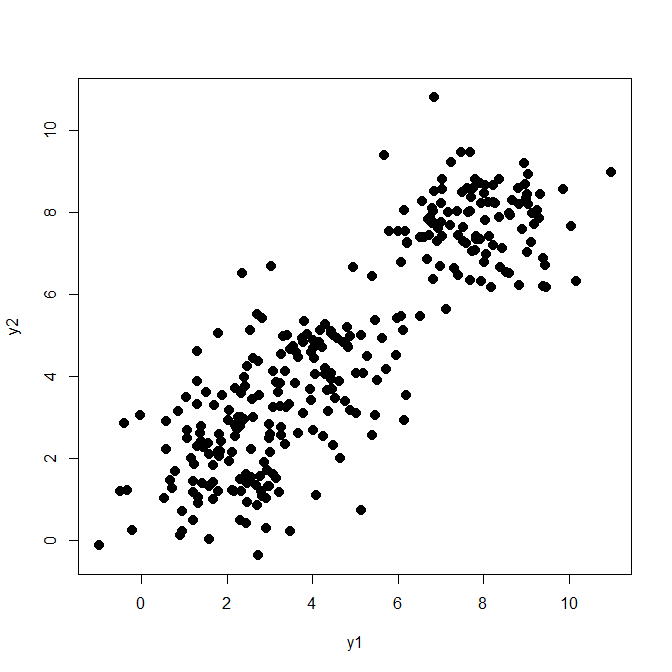}
	\includegraphics[width=0.325\textwidth, trim = 0 20 0 50, clip]{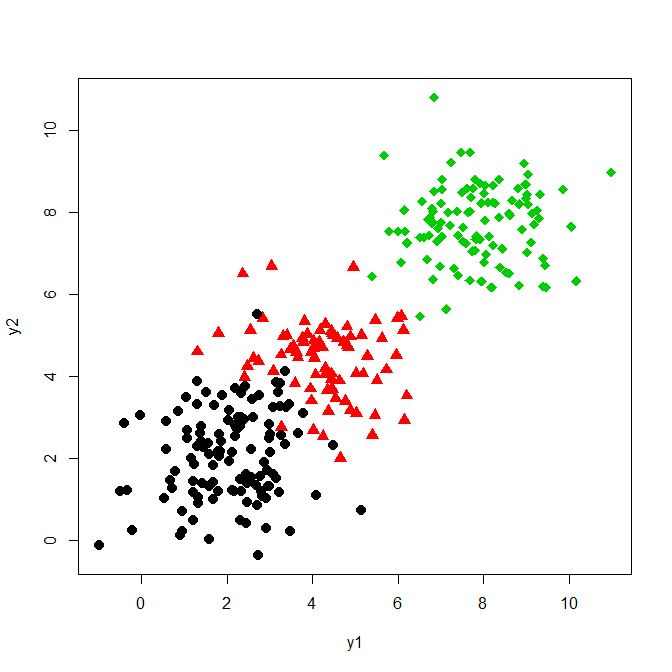}
	\includegraphics[width=0.325\textwidth, trim = 0 20 0 50, clip]{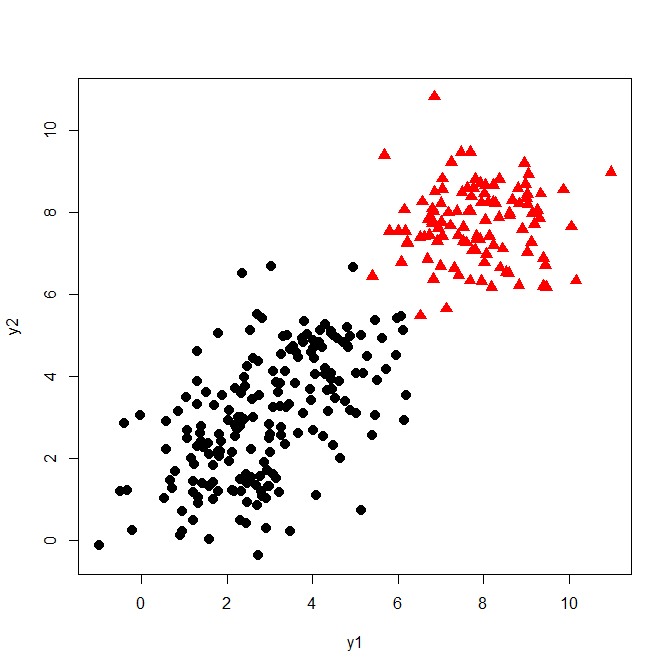}
	\caption{\footnotesize Simulation setup II. Scatter plot of one
		data set (left-hand side), classification according to the
		generating distributions (middle) and to the clusters
		\comment{obtained from a mixture of mixtures with $K=10$, $L=4$, $\phi_B=0.5$ and $\phi_W=0.1$}
		(right-hand side).}\label{plot:3points}
\end{figure}

In order to mimic close data clusters, \comment{10 data sets} with 300 observations
are generated from a three-component normal mixture, where, however,
only two data clusters can be clearly distinguished. In Figure
\ref{plot:3points} the scatter plot of one data set is displayed. The
300 observations are
sampled from a normal mixture with component means
\begin{equation*}\label{}
(\bmu_1 \quad \bmu_2 \quad \bmu_3)=  \begin{pmatrix} 2 &4.2&7.8\\ 2&4.2&7.8\end{pmatrix},
\end{equation*}
variance-covariance matrices $\bSigma_1=\bSigma_2=\bSigma_3=\bI_2$ and
equal weights $\boldeta=(1/3,1/3,1/3)$.

For {\bl various} values of
$\phi_B$ (between 0.1 and 0.9) and $\phi_W$ (between 0.01 and 0.4) a
sparse  mixture of mixtures model with $K=10$ clusters and $L=4$
subcomponents is fitted and the number of clusters is
estimated. For each combination of $\phi_B$ and $\phi_W$ the results
are reported in Table~\ref{table:3points}.

Table~\ref{table:3points} indicates that if $\phi_B$ increases, also
$\phi_W$ has to increase in order to identify exactly two
clusters. This makes sense since by increasing $\phi_B$ the a~priori
within-cluster variability becomes smaller yielding tight subcomponent
densities.  Tight subcomponents in turn require a large proportion
$\phi_W$ of variability explained by the subcomponent means to capture
the whole cluster. Thus $\phi_W$ has to be increased too. However,
$\phi_W$ has to be selected carefully. If $\phi_W$ is larger than
actually needed, some subcomponents are likely to ``emigrate''
\commentG{to  neighboring} clusters. This leads finally to only one cluster
being estimated for some data sets. This is basically the case for
some of the combinations of $\phi_B$ and $\phi_W$ displayed in the
upper triangle of the table.  In contrast, if $\phi_W$ is smaller than
needed, due to the induced shrinkage of the subcomponent means toward
the cluster center, the specified cluster mixture distribution is not
able to fit the whole data cluster and two cluster distributions are
needed to fit a single data cluster. This can be seen for some of the
combinations of $\phi_B$ and $\phi_W$ displayed in the lower triangle
of the table.

\begin{table}[t!]
	\centering
	\begin{footnotesize}
		\begin{tabular}{|c|c|c|c|c|c|}
			\hline
			\backslashbox{$\phi_B$}{$\phi_W$}&0.01   & 0.1   & 0.2  &  0.3 & 0.4  \\ \hline
			0.1 & 3(6)&{2}(10)&2(5)&1(8)& 1(8) \\
			& 2(4)&     &1(5)&2(2)& 2(2) \\   \hline
			0.3  & 3(6)&{2}(10)&2(8)&2(6)& 1(7) \\
			& 2(4)&    &1(2)&1(4)& 2(3) \\  \hline
			0.5  & 3(5)&{2}(10)&{2}(10)&2(9)&2(7) \\
			& 2(5)&    &    &1(1)&1(3) \\  \hline
			0.7  & 3(7)&2(7)&{2}(10)&{2}(10)&{2}(10) \\
			& 2(3)&3(3)&    &   &\\  \hline
			0.9  & 3(6)&3(7)&3(5)&2(8)&{2}(10) \\
			& 4(4)&2(3)&2(5)&3(2)&     \\
			\hline
		\end{tabular}
	\end{footnotesize}
	\caption{\comment{\footnotesize Simulation setup II (based on 10 data sets); true number of clusters equal to 2, $K=10$, $L=4$.   Results for the estimated number of non-empty clusters $\hat{K}_0$ for {\bl various} amounts of $\phi_B$ and $\phi_W$.   The number of data sets estimating the reported  $\hat{K}_0$
			is given in parentheses.}} \label{table:3points}
	%\footnotesize Simulation setup II, number of data sets $=10$, $K=10$, $L=4$. Simulation results  for estimating the number of clusters ($\hat{K}_0$) for different amounts of $\phi_B$ and $\phi_W$.  The number of data sets estimating the reported $\hat{K}_0$ is given in parentheses.}\label{table:3points}
\end{table}

\section{Description of the data sets}\label{app:datasets}

The following data sets are investigated.  The Yeast data set \citep{Mix:NakaiKanehisa1991} aims at
predicting the cellular localization sites of proteins and
%It has been previously been studied by \cite{Mix:NakaiKanehisa1991} and
can be
downloaded from the UCI machine learning repository
\citep{mix:Bache+Lichman:2013}. As in
\cite{Mix:FranczakBrowneMcNicholas2012}, we aim at distinguishing
between the two localization sites CYT (cytosolic or cytoskeletal) and
ME3 (membrane protein, no N-terminal signal) by considering a subset
of three variables, namely McGeoch's method for signal sequence (mcg),
the score of the ALOM membrane spanning region prediction program
(alm) and the score of discriminant analysis of the amino acid content
of vacuolar and extracellular proteins (vac).

The Flea beetles data set \citep{Mix:Lubischew1962} considers 6
physical measurements of 74 male flea beetles belonging to three
different species. It is available in the \proglang{R} package
\pkg{DPpackage} \citep{Mix:JaraHansonQuitanaMuellerRosner2011}.

The Australian Institute of Sport (AIS) data set
\citep{Mix:CookWeisberg1994} consists of 11 physical measurements on
202 athletes (100 female and 102 male).  As in
\cite{Mix:LeeMcLachlan2013}, we only consider three variables, namely
body mass index (BMI), lean body mass (LBM) and the percentage of body
fat (Bfat). The data set is contained in the \proglang{R} package
\pkg{locfit} \citep{Mix:Loader2007}.

The Breast Cancer Wisconsin (Diagnostic) data set
\citep{Mix:MangasarianStreetWolberg1995} describes characteristics of
the cell nuclei present in images. The clustering aim is to
distinguish between benign and malignant tumors.  It can be downloaded
from the UCI machine learning repository.  Following
\cite{Mix:FraleyRaftery2002} and \cite{Mix:Viroli2010} we use a subset
of three attributes: extreme area, extreme smoothness, and mean
texture. Additionally, we scaled the data.

\begin{figure}[t!]
	\centering
	\includegraphics[width=0.325\textwidth]{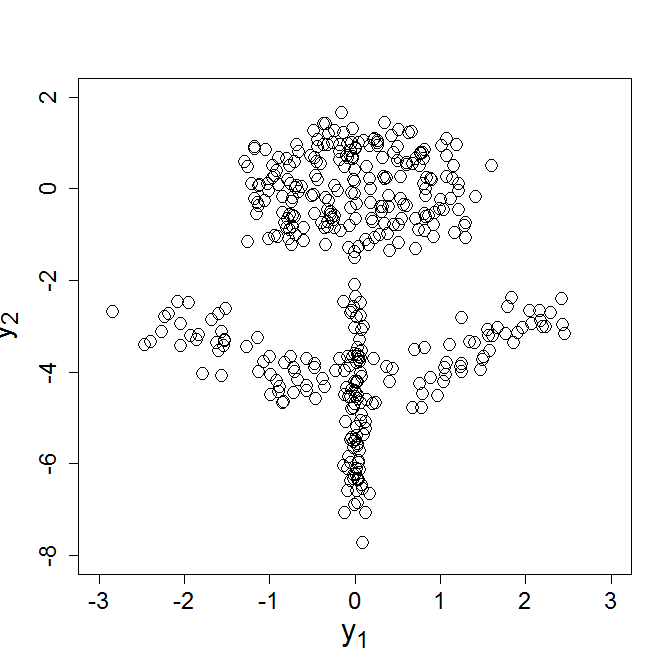}
	\includegraphics[width=0.325\textwidth]{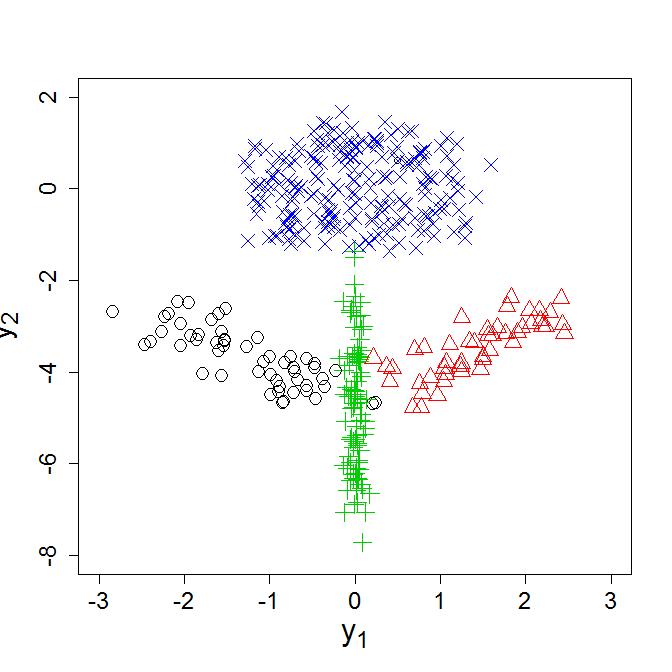}
	\includegraphics[width=0.325\textwidth]{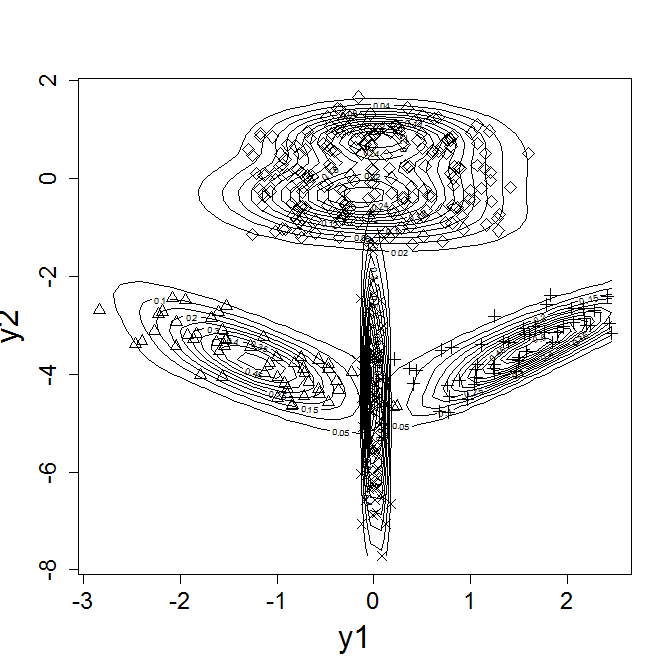}
	\caption{\footnotesize Flower data set. Boxplot of a sample with
		400 data points (left-hand side), the estimated clusters for
		$K=10, L=4$, $\phi_B=0.5,\phi_W=0.1, \nu_1=\nu_2=10$ (middle),
		and the corresponding cluster distributions (right-hand
		side).} \label{plot:flower}
\end{figure}

The artificial flower data set reported by \cite{Mix:Yerebakan2014}
can be downloaded from \url{https://github.com/halidziya/I2GMM}. It
consists of {\bl $17000$} two-dimensional observations representing a
flower shape. The data set is generated by seventeen Gaussian
densities forming 4 clusters: nine components generate the blossom of
the flower, four components the stem and two components each of the
two leaves.  Note that within each cluster, the generating components
have the same orientation. This specification meets the assumption
made in the infinite mixture of infinite mixtures model by
\cite{Mix:Yerebakan2014}.  We used a subsample of 400 data points for
our application, thus leading to the benchmark data sets all being of
comparable size. The scatter plot of the sample is given in
Figure~\ref{plot:flower} on the left-hand side. If we fit a sparse
mixture of mixtures model with $K=10$ clusters and $L=4$ subcomponents
and the usual prior settings as described in Section 2, the four
clusters of the flower (petal, stem, and two leaves) can be clearly
captured, as can be seen in Figure~\ref{plot:flower}, where the
estimated clustering result and the corresponding cluster
distributions are shown.

The flow cytometry data set DLBCL
%To demonstrate the application , we consider the clustering of a three-dimensonal Diffuse Large B-cell Lymphoma (DLBCL) dataset.\\
contains intensities of markers stained on a sample of over 8000
cells derived from the lymph nodes of patients diagnosed with Diffuse
Large B-cell Lymphoma (DLBCL).  The aim of the clustering is to group
the individual cell data measurements into only a few groups on the
basis of similarities in light scattering and fluorescence, see
\cite{Mix:Aghaeepour2013} for more details. For this data set, class
labels of the observations {\r partitioning the data into four
	classes are available which were} obtained by manual partitioning
(``gating'').  The data set is available in the \proglang{R} package
\pkg{EMMIXuskew} \citep{Mix:LeeMcLachlan2013b} as data set
\texttt{DLBCL} with the corresponding class labels in
\texttt{true.clusters}.

The flow cytometry data set \emph{GvHDB01case} by
\cite{Mix:Brinkman2007} consists of {\bl 12442} six-dimensional observations
which represent a blood sample from a subject who developed Graft
versus Host disease (GvHD). GvHD is a severe complication following a
blood and marrow transplantation, when donor immune cells in the graft
attack the body cells of the recipient.
The data were analyzed first by \cite{Mix:Brinkman2007}. % and then,
%among others, by \cite{Mix:FruehwirthPyne2010} using mixtures.
\cite{Mix:Lo2008} fitted a Student-$t$ mixture model to this
data and estimated 12 clusters using the EM algorithm. In the Bayesian
framework,  \cite{Mix:FruehwirthPyne2010} fitted finite mixtures of
skew-normal and skew-$t$ distributions and found 12 and 9 clusters.
By comparing this sample to a
control sample from a patient who had a similar transplantation but
did not develop the disease, \cite{Mix:Brinkman2007} found a very
small cluster of live cells (high ``FSC'', high ``SSC'') in the sample
with a high expression in the four markers (``CD4+'',
``CD8$\beta$+''', ``CD3+'', ``CD8+''). This cluster was not present in
the control sample and seems to be correlated with the development of
GvHD.

For the data sets with known class labels, the clustering result of
the estimated models is measured by the misclassification rate and
the adjusted Rand index \citep{Mix:HubertArabie1985}.  To calculate
the misclassification rate of the estimated model, the ``optimal''
matching between the estimated cluster labels and the true known class
labels is determined as the one minimizing the misclassification rate
over all possible matches for each of the scenarios.  The
misclassification rate is measured by the number of misclassified
observations divided by all observations and should be as small as
possible.

The adjusted Rand index \citep{Mix:HubertArabie1985} is used to assess
the similarity between the true and the estimated partition of the
data.  It is a corrected form of the Rand index \citep{Mix:Rand1971}
which is adjusted for chance agreement. An adjusted Rand index of 1
corresponds to perfect agreement of two partitions whereas an adjusted
Rand index of 0 corresponds to results no better than would be
expected by randomly drawing two partitions, each with a fixed number
of clusters and a fixed number of elements in each cluster.

%The Old Faithful Geyser data set \citep{mix:azzalinibowman1990} is
%available as data set \texttt{faithful} in the \proglang{R} package
%\pkg{datasets} \citep{Mix:R2012}, It measures the waiting time between
%eruptions and the duration of the eruption for the Old Faithful geyser
%in Yellowstone National Park, Wyoming, USA. There exists no known
%classification for this data set.

\section{Issues with the merging approach}

\begin{figure}[t!]
	\centering
	\includegraphics[width=0.32\textwidth]{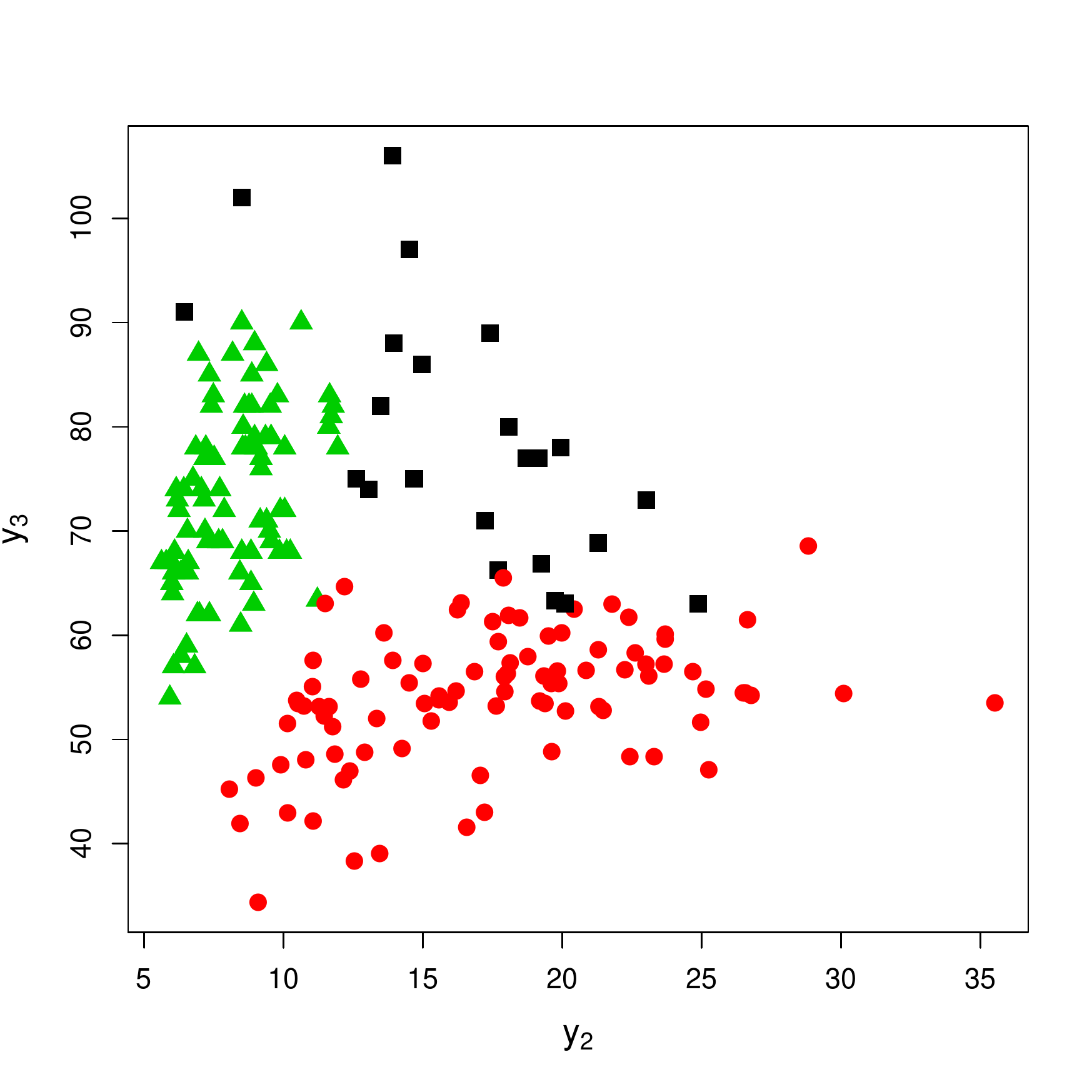}
	\includegraphics[width=0.32\textwidth]{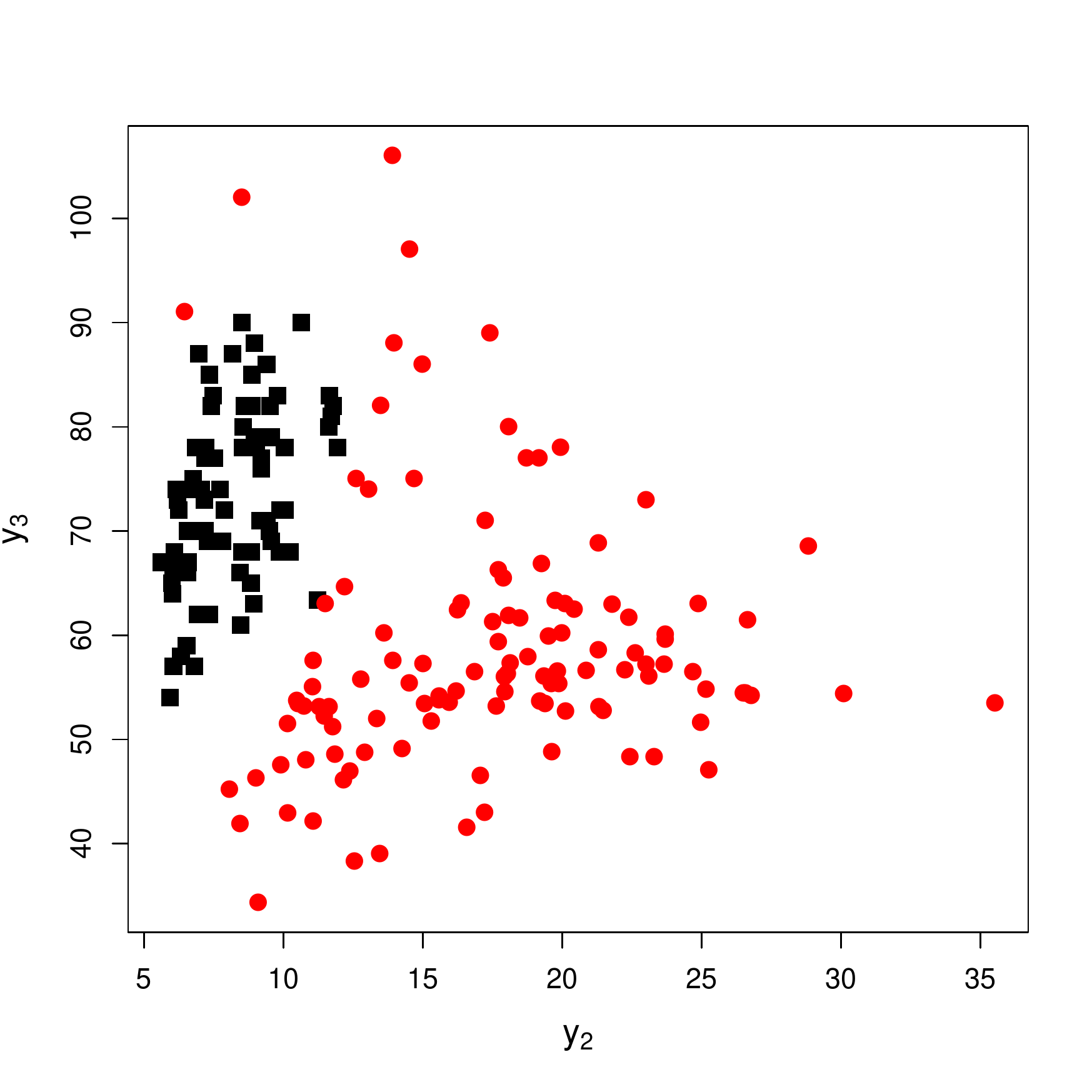}
	\includegraphics[width=0.32\textwidth]{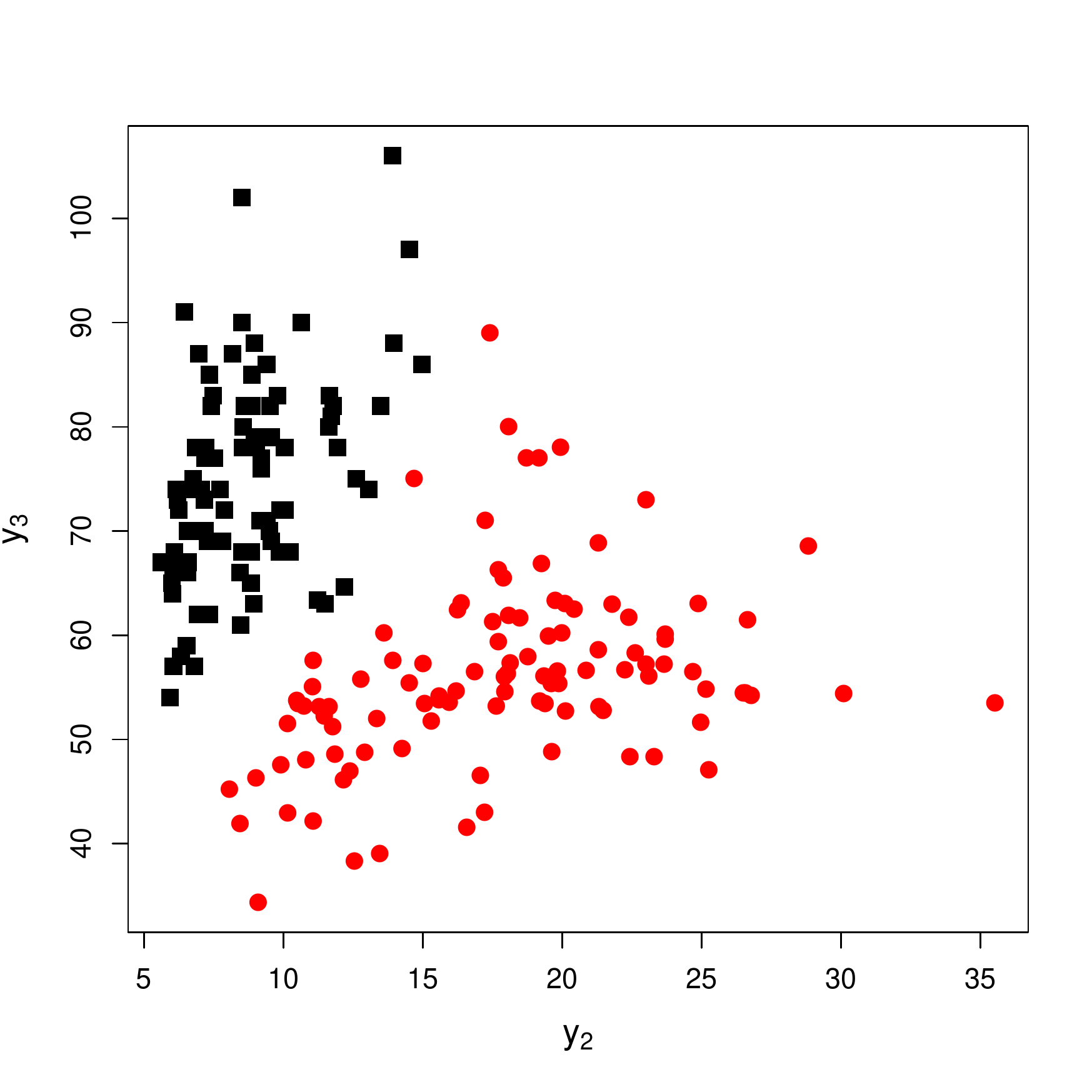}
	\caption{\footnotesize AIS data set, variables ``X.Bfat'' and
		``LBM''. Scatter plots of the observations with different
		estimated classifications based on
		\texttt{Mclust} % $\hat{K}_0=3$
		(left-hand side),  \texttt{combiClust} (middle), % $\hat{K}_0=2$
		%Bottom: Classification estimated by a sparse finite
		% mixture, $K=10,L=1$, $\hat{K}_0=3$ (left-hand side)
		and the  sparse hierarchical mixture of mixtures approach developed in this paper ($K=10, L=4$) (right-hand
		side).}\label{plot:AisR3}
\end{figure}

The merging approach, which consists of first fitting a finite mixture
of Gaussians to suitably approximate the data distribution and
subsequently combines components to clusters, is susceptible to yield
poor classifications, since the resulting clusters can only emerge as
the union of components that have been identified in the previous
step. %under the misspecified Gaussian assumption. % in the first place.
For illustration, the AIS data (see Appendix~D) are clustered using
function \texttt{clustCombi}
\citep{Mix:BaudryRafteryCeleuxLoGottardo2010} from the \proglang{R}
package \pkg{mclust} \citep{Mix:FraleyRafteryScrucca2012}.  The
results are shown in Figure~\ref{plot:AisR3}. The first step results
in a standard Gaussian mixture with three components (left-hand plot),
and subsequently \emph{all data} in the smallest component are merged
with one of the bigger components to form two clusters (middle plot)
which are not satisfactorily separated from each other due to the
misspecification of the standard Gaussian mixture in the first step.
In contrast, the sparse hierarchical mixture of mixtures approach we
develop in the present paper identifies two well-separated clusters on
the upper level of the hierarchy (right-hand plot).

\section{Fitting a mixture of two $\mathit{SAL}$ distributions}\label{app:SAL}
\begin{figure}[t!]
	\centering
	\includegraphics[width=0.325\textwidth]{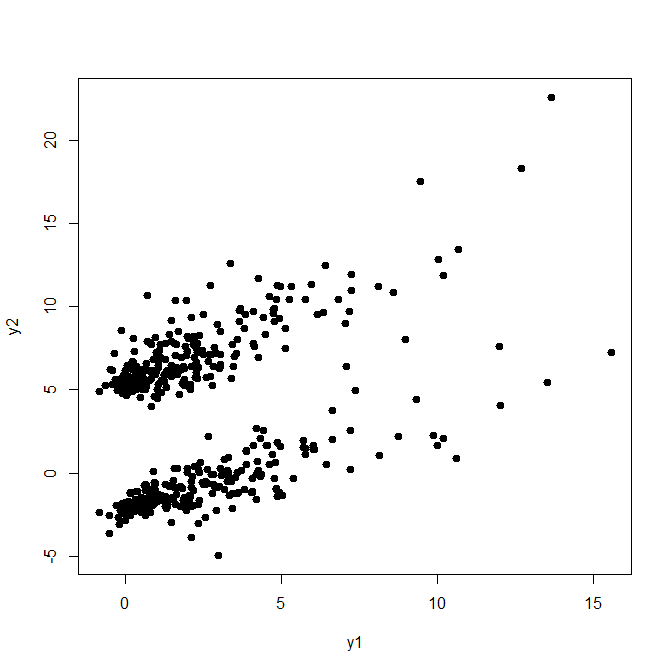}
	\includegraphics[width=0.325\textwidth]{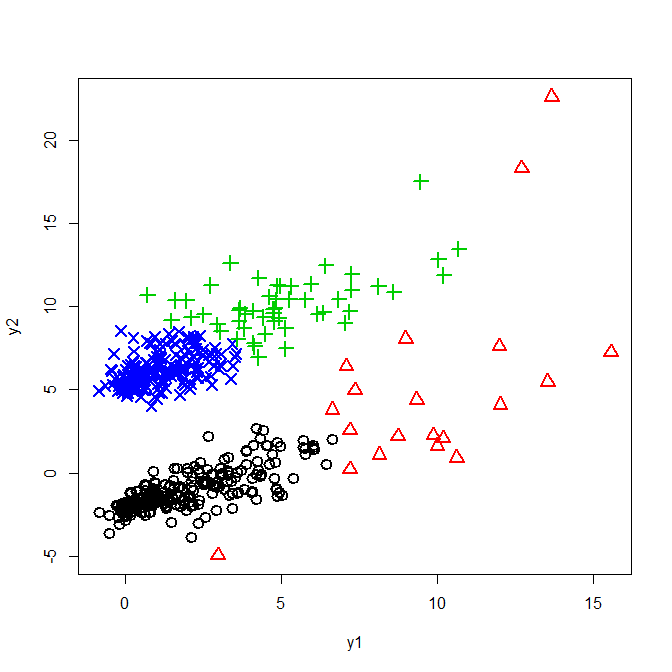}
	\includegraphics[width=0.325\textwidth]{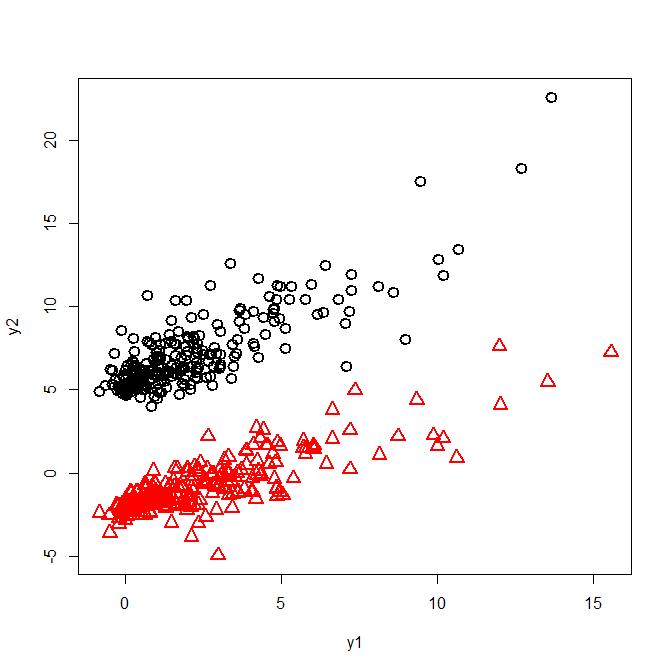}
	\caption{\footnotesize Samples from a mixture of two $\mathit{SAL}$ distributions
		(left-hand side), the estimated clusters for $K=10, L=4$,
		$\phi_B=0.5,\phi_W=0.1, \nu_1=\nu_2=10$ (middle), and for $K=10,
		L=5$, $\phi_B=0.4,\phi_W=0.2$, with fixed hyperparameters
		\comment{$\bC_{0k}=g_0 \cdot\bG_0^{-1}$ and $\lambda_{kj} \equiv 1$} %$\bC_{0k}$ and $\lambda_{kl}$
		(right-hand
		side).} \label{plot:SalData}
\end{figure}

Although it is not the purpose of our approach to capture non-dense
data clusters, we apply it to the challenging cluster shapes generated
by \comment{shifted asymmetric Laplace ($\mathit{SAL}$)}  distributions, which are introduced by
\cite{Mix:FranczakBrowneMcNicholas2012} in order to capture asymmetric
data clusters with outliers.  We sampled data from a mixture of two
$\mathit{SAL}$ distributions according to Section~4.2 in
\cite{Mix:FranczakBrowneMcNicholas2012}. The data set is shown in
Figure~\ref{plot:SalData} on the left-hand side.

If we fit a sparse hierarchical mixture of mixtures model with $K=10$
clusters and $L=4$ subcomponents and priors and hyperpriors specified
as in \comment{Sections~2.1~and~2.3}, four clusters are estimated, as can be seen in the
middle plot of Figure~\ref{plot:SalData}. Evidently, the standard
prior setting, tuned to capture dense homogeneous data clusters,
performs badly for this kind of clusters. Thus, in order to take the
specific data cluster shapes into account, we adjust the prior
specifications accordingly. A data cluster generated by a
$\mathit{SAL}$ distribution is not homogeneously dense, it rather
consists of a relatively dense kernel on one side of the cluster and a
non-dense, light and comet-like tail with possibly extreme
observations on the other side. Therefore within a cluster,
subcomponents with very different covariance matrices are required in
order to fit the whole cluster distribution. Since specification of
hyperpriors on $\lambda_{kj}$ and $\bC_{0k}$ has a smoothing and
balancing effect on the subcomponent densities, we omit these
hyperprior specifications, and choose fixed values for $k=1,\ldots,K$,
i.e. $\bC_{0k}=g_0 \cdot\bG_0^{-1}$ and $\lambda_{kj} \equiv 1$,
$j=1,\ldots, r$.

Additionally, in
order to reach also extreme points, we increase both the number of
subcomponents to $L=5$ and the a~priori variability explained by the
subcomponent means to $\phi_W=0.2$. At the same time we adjust the
proportion of heterogeneity explained by the cluster means by
decreasing $\phi_B$ to $0.4$, thus keeping the subcomponent covariance
matrices large. If we estimate again a sparse hierarchical mixture of mixtures
model with these modified prior settings, the two clusters can be
identified, see Figure~\ref{plot:SalData} on the right-hand side.

\bibliographystyle{Chicago}
\bibliography{MixofMix}

\end{document}

%% file: newcommands.tex
\newcommand{\by}{\mathbf{y}}

\newcommand{\bm}{\mathbf{m}}

\newcommand{\bw}{\mathbf{w}}

\newcommand{\bY}{\mathbf{Y}}
\newcommand{\bb}{\mathbf{b}}
\newcommand{\bB}{\mathbf{B}}
\newcommand{\bC}{\mathbf{C}}
\newcommand{\bG}{\mathbf{G}}
\newcommand{\bM}{\mathbf{M}}

\newcommand{\bS}{\mathbf{S}}

\newcommand{\bI}{\mathbf{I}}

\newcommand{\btheta}{\boldsymbol{\theta}}
\newcommand{\bTheta}{\boldsymbol{\Theta}}

\newcommand{\boldeta}{\boldsymbol{\eta}}
\newcommand{\bmu}{\boldsymbol{\mu}}

\newcommand{\bSigma}{\boldsymbol{\Sigma}}

\newcommand{\bLambda}{\boldsymbol{\Lambda}}

\newcommand{\cN}{\mathcal{N}}
\newcommand{\cG}{\mathcal{G}}

\newcommand{\cW}{\mathcal{W}}

\newcommand{\stick}{v}
\newcommand{\stickw}{u}
\newcommand{\iid}{\mbox{\rm i.i.d.}}
\newcommand{\Betadis}[1]{\mathcal{B}\left(#1\right)}
\newcommand{\ym}{{\mathbf y}} % y multivariate observation
\newcommand{\Cl}{C}
\newcommand{\parti}{{\cal P}}
\newcommand{\Kn}{K _0}
\newcommand{\Kni}{\Kn ^{-i}}
\newcommand{\Nki}[1]{N_{#1} ^{-i}}
\newcommand{\ed}[1]{e_{#1}} % Dirichlet prior
\newcommand{\Siv}{{\mathbf{S}}} % indicatoren
\newcommand{\etav}{{\boldsymbol{\eta}}} % vector of weights
\newcommand{\Gamfun}[1]{\Gamma (#1)}
\newcommand{\Count}[1]{\#\{#1\}}
\newcommand{\Prob}[1]{\mbox{\rm Pr}\{#1\}}
\newcommand{\Dirinv}[2]{\mathcal{D}_{#1}\left(#2\right)}
\newcommand{\alphaDP}{\alpha}
\newcommand{\alphaPY}{\alpha}
\newcommand{\betaPY}{\beta}
\newcommand{\hyp}{\phi_0}

\newcommand{\simiid}{\overset{iid}{\sim}}